\documentclass[10pt,prd,twocolumn,showpacs,amsmath,amssymb,nofootinbib,preprintnumbers,superscriptaddress]{revtex4-1}
\usepackage{wasysym}
\usepackage{graphicx,color}
\usepackage{dcolumn}
\usepackage{bm}
\usepackage{amssymb}
\usepackage{epsfig}
\usepackage{slashed}
\usepackage{hyperref}
\usepackage{graphicx,amsmath,amssymb,bm}
\newcommand{\be}{\begin{equation}}
\newcommand{\ee}{\end{equation}}
\newcommand\beq{\begin{eqnarray}}
\newcommand\eeq{\end{eqnarray}}

\newcommand{\CO}{{\cal O}}

\begin{document}

\preprint{YITP-SB-14-11}

 \title{The holographic dual of a Hawking pair has a wormhole}
\author{Kristan Jensen}
\email[Email: ]{kristanj@insti.physics.sunysb.edu}
\affiliation{C. N. Yang Institute for Theoretical Physics, SUNY, Stony Brook, NY 11794-3840, USA}
\author{Andreas Karch}
\email[Email: ]{akarch@uw.edu}
\affiliation{Department of Physics, University of Washington, Seattle, WA
98195-1560, USA}
\author{Brandon Robinson}
\email[Email: ]{robinb22@uw.edu}
\affiliation{Department of Physics, University of Washington, Seattle, WA
98195-1560, USA}

\begin{abstract}
In previous work, two of us constructed the holographic dual of a single Einstein-Podolsky-Rosen (EPR) pair in maximally supersymmetric Yang-Mills theory. Here we extend that work in two directions. First, we use a Randall-Sundrum brane to couple the entangled pair to dynamical gravity. Second, we consider the entangled pair in the background of a black hole with the particles on opposite sides of the horizon, turning the EPR pair into a Hawking pair. We give general arguments that ER=EPR should be understood as a duality between two equivalent descriptions of the same physical reality.
\end{abstract}

\date\today

\maketitle

\section{Introduction}

Entanglement is correlation. While it is easy to write down the state vector for an entangled pair of spins, the state itself is not measurable. Rather, the observable consequence of entanglement is non-trivial correlation between the spins. If the system under consideration can be split into two subsystems with operators acting only on subsystem A or B respectively, it follows from the basic principles of quantum mechanics that, in a pure product state $|ab\rangle=|a\rangle\otimes|b\rangle$,
\be
	\langle a b|  O_A O_B | a b\rangle = \langle a | O_A |a \rangle \langle b |O_B | b \rangle\,.
\ee
Here $|a \rangle$ and $| b \rangle$ are states describing subsystems A and B respectively, and $O_A$ and $O_B$ are operators only acting on the corresponding subsystem. The connected equal-time two-point functions
\be
\label{connected}
	G_{AB} = \langle \psi|  O_A O_B | \psi \rangle - \langle \psi | O_A | \psi \rangle \langle \psi |O_B | \psi \rangle\,,
\ee
indicate entanglement, in the sense that $G_{AB}$ vanishes in product states. More precisely, by causality $G_{AB}\neq 0$ cannot arise from interactions between degrees of freedom in A and B and so indicates non-trivial correlations in the density matrix describing the system. When the system is in a pure state, $G_{AB}$ quantifies entanglement. In the example of an Einstein-Podolsky-Rosen (EPR) pair, where A and B stand for two spatially separated spins that we put into the maximally entangled pure state given by
\be
	| \psi \rangle = \frac{1}{\sqrt{2}} \left (| \uparrow \downarrow \rangle - | \downarrow \uparrow \rangle \right )\,,
\ee
 we have
\be
	\langle \vec{S}_A \rangle = \langle \vec{S}_B \rangle =0\,,
\ee
but
\be
\label{E:spin2pt}
	\langle (\hat{n}\cdot\vec{S}_A) (\hat{m}\cdot  \vec{S}_B) \rangle = - \frac{\hat{n}\cdot\hat{m}}{4}\,.
\ee
All one-particle spin measurements are completely undetermined. Thus, all of the information about the quantum state is in the correlations. When measuring any component of spin on both particles, the likelihood for them to point in opposite directions is 100\%.

For any local quantum field theory, we can associate the subsystems A and B with spatially separated regions of spacetime. Any local operator which acts inside the causal development of region A is denoted $O_A$, and similarly for $O_B$. If the connected two-point function $G_{AB}$ is nonzero, then the system is in a global state where local states in region A and B are correlated with one another. Note that this is the case for the vacuum of standard quantum field theories. In this sense entanglement is directly related to connectedness of the space, as pointed out in \cite{VanRaamsdonk:2010pw}.

Going back to the EPR pair, what we really have in mind is the following scenario: create the entangled spins at some initial time and then separate them in such a way that they are out of causal contact forever after. Consequently, no local spin-spin interactions can decohere the correlations~\eqref{E:spin2pt}, as measured at later times. While the entangled spins are out of causal contact, they are still ``connected" in the sense of having connected correlations at spacelike separation. Thus, we are discussing entanglement between causally disconnected degrees of freedom, and it is this sort of entanglement that will be the primary subject of this work. This is somewhat different than the entanglement between two spatially separated regions of space in local QFT. If our theory lives in flat Minkowski space, then degrees of freedom in any spatial regions A and B can eventually communicate provided that we are sufficiently patient. The ``EPR entanglement'' is more akin to putting a local QFT on a manifold with several disconnected components and preparing a state in which degrees of freedom on different components are entangled with each other.	

In theories with a holographic dual this statement carries even more meaning. Holography asserts that a quantum system in $d$ spacetime dimensions has a dual description in terms of quantum gravity in $d+1$ dimensions \cite{Maldacena:1997re,Gubser:1998bc,Witten:1998qj}, where one may think of the $d$-dimensional field theory as living on the boundary of the bulk manifold. In the limit that the gravitational theory is well approximated by classical gravity, to compute the two-point function for a particular local operator one calculates the classical Green's function for the field dual to this operator. If the system was in a product state with no entanglement whatsoever, then all such correlations, and so all bulk classical propagators connecting the operator insertions, must vanish. This, in turn, would imply that the spatial distance between the two local operators in the holographic bulk theory is infinite. As long as the two points are at all connected in the bulk, there will be some non-vanishing Green's functions. In this sense, the entanglement in the field theory encodes the connectedness of the bulk manifold.

This statement becomes particularly interesting when applied to an EPR pair, or EPR entanglement between two causally disconnected regions A and B in the field theory.  In the gravity dual, there are asymptotically AdS regions with boundaries A and B which are separated by a finite spatial distance, while at the same time shielded against causal communication. If, as in the case of two Rindler wedges of flat space, A and B are separated by causal horizons in the field theory, then bulk causality is implemented by causal horizons between the causal developments of A and B into the bulk. However, if A and B are not separated by causal horizons in the field theory, then they are separated by event horizons. Either way, these are the defining features of a non-traversable wormhole, or Einstein-Rosen (ER) bridge. These simple arguments imply that whenever we are describing a field theory state in which causally disconnected regions are entangled with each other, the gravity dual has to contain a non-traversable wormhole.\footnote{We caution that the gravity dual of a general entangled state need not be a wormhole geometry which is everywhere smooth and weakly curved~\cite{Marolf:2013dba}. For instance, consider a state in which only a subset of operators $\mathcal{O}_i$ exhibit large EPR correlations. Presumably the gravity dual is out of the semiclassical supergravity approximation.} We explicitly demonstrated this for an EPR pair made of a heavy quark-antiquark pair in maximally supersymmetric Yang-Mills (mSYM) theory in \cite{Jensen:2013ora}. The holographic dual of EPR (entanglement between causally disconnected regions) is ER. This equivalence follows directly from the basic rules of holography for the calculation of correlation functions.

Our investigations into the relation between entanglement and wormholes were, of course, strongly motivated by Maldacena and Susskind's recent proposal that ER=EPR \cite{Maldacena:2013xja}. In their work, they argue that ER is not just the holographic dual of EPR, but rather, due to quantum gravitational effects any entangled pair is connected by a wormhole in the same spacetime. It is difficult to precisely formulate what this statement implies. For instance, the simplest operational definition of entanglement is the existence of correlations between local operators in causally disconnected regions. However in quantum gravity, there are no gauge-invariant local operators, and so the correlations defining entanglement are not good observables. Although, in macroscopic spacetimes with a good semi-classical description, the concept of local observables should make sense and one can ask whether, in this context, ER=EPR.

In this work, we take the first steps toward connecting our statement that ``EPR is holographically dual to ER" with the Maldacena-Susskind conjecture that ``ER=EPR". First, we use the Randall-Sundrum construction \cite{Randall:1999vf} to couple mSYM to dynamical gravity. We will show that the holographic dual of an EPR pair in a world governed by quantum mechanics coupled to gravity is still a geometry with a non-traversable wormhole. Second, we use the freedom in the holographic duality to change the background spacetime on which mSYM is formulated in order to study an EPR pair in mSYM on a black hole background. In particular, we show that the holographic dual for an EPR pair with the two particles on opposite sides of the black hole horizon (which we will refer to as a Hawking pair) has an ER bridge. Last but not least, we bring these two modifications together to demonstrate that a Hawking pair in a theory with dynamical gravity and black holes has an equivalent description in terms of classical gravity with an ER bridge. That is, the effects of entanglement in a quantum theory coupled to gravity can be encoded in a classical wormhole geometry.

While still not quite the ER=EPR statement of \cite{Maldacena:2013xja}, we hope that this holographic equivalence correctly captures the essence of the ER=EPR proposal. Perhaps this is the way ER=EPR should have originally been formulated. To wit, the primary examples used in \cite{Maldacena:2013xja} to argue for ER=EPR involve entangled pairs of black holes whose exteriors were shown to be connected by an ER bridge. An isolated black hole can be thought of as either a classical solution of general relativity (GR) with an entropy $S$ proportional to the horizon area and energy $E$ or in terms of is microscopic constituents, {\it{e.g.}} D-branes or strings, which give rise to an $e^S$-fold degeneracy of states with the same energy. In principle, the microscopic description exists for any number of constituents.  However, only when the effective coupling constant is small can the microscopic description be treated perturbatively, while the black hole description is perturbative only in the large entropy and large coupling limit. In this sense the black hole geometry is emergent at large $S$, but we can define what we mean by a black hole of any size through the microscopic ensemble. Similarly, one presumably should conclude that we can think of the entangled black hole pair as a classical wormhole geometry in the large area limit or as an entangled state of the two sets of microscopic degrees of freedom at any scale. Just as for the emergent black hole geometry, these should be complementary descriptions that are not simultaneously tractable. This is the description suggested by our holographic analysis.

The remainder of this article is organized as follows. In Section~\ref{S:RS} we review the Randall-Sundrum (RS) scenario in the context of holography, which we then use to study entangled pairs in mSYM coupled to dynamical gravity. Next in Sections~\ref{S:Hawking} and~\ref{S:asyFlat} we study mSYM and entangled pairs in two-sided black hole backgrounds. We focus on two principle examples of entangled pairs: (i.) we consider entangled pairs for mSYM on a planar AdS black brane, where one quark is outside one horizon and the anti-quark is outside the other horizon, and (ii.) the corresponding pairs for mSYM on a Schwarzschild black hole. In this sense these entangled pairs are Hawking pairs. We put these two ideas together in Section~\ref{S:HawkRS}, looking at Hawking pairs in the RS scenario. In our penultimate Section~\ref{S:cosmo} we consider entangled pairs in backgrounds with cosmological horizons, and thereafter conclude in Section~\ref{S:conclude}.

\section{Entangled Pairs in Randall-Sundrum models}
\label{S:RS}

\subsection{Holographic duals for mSYM coupled to 4d gravity}

The simplest holographic dualities relate gravity in AdS$_{d+1}$ to a conformal field theory (CFT) in $d$ dimensions. For concreteness throughout the article, we will focus on the $d=4$ case and the equivalence between type IIB string theory on AdS$_5\times \mathbb{S}^5$ and mSYM theory, although in many cases we present formulae for general $d$. Also throughout, we will work in the usual Maldacena limits such that the mSYM is strongly coupled and the string dual is classical. The Randall-Sundrum (RS) scenario \cite{Randall:1999vf} introduces one more ingredient into this setup: a codimension-one brane of tension $T_{RS}$ described by a minimal area action. Since the gravitational backreaction of the brane is delta function localized, the spacetime is still AdS$_{d+1}$ away from the brane. We write the $d+1$-dimensional metric as
\be
\label{slicing}
	g = e^{2 A(z)} \left ( g_{d} + dz^2 \right )\,.
\ee
Without loss of generality we put the RS brane at a constant $z=0$ so that the induced metric on the brane is proportional to the $d$-dimensional metric $g_d$ (which we take to be independent of $z$). Einstein's equations then enforce two important properties: i) $g_d$ must be an Einstein metric and ii) the brane cuts off the original space at a constant value $z_*$ and glues it together with a second copy of the same spacetime. This can be encoded in a change of the warpfactor after including the RS brane,
\be
\label{rs} A(z) \rightarrow A(|z| + z^*)\,.
\ee
The metric is continuous across the cut, but its derivatives are not. The corresponding jump in extrinsic curvature $K_{ij}$ is supported by the delta function source as encoded in the Israel junction conditions \cite{Israel:1966rt}
\be
\label{E:junction}
	\lim_{\epsilon \to 0}
	\left . ( K_{ij} - g_{ij} K ) \right |_{-\epsilon}^{+ \epsilon} = 8 \pi G_N t_{ij} \,,
\ee
where $G_N$ is the $d+1$ dimensional Newton's constant and $t_{ij}$ the stress tensor of the brane, {\it{i.e.}} $t_{ij}=2/\sqrt{-h}\left(\delta S/\delta h_{ij}\right)$ where $h_{ij}$ is the induced metric.
In addition to relating the tension of the RS-brane to the warp factor $A(z)$, the junction condition enforces that $z_*$ is positive when the RS-brane has positive tension.

Since $g_d$ must be Einstein, there are three broad classes of RS scenarios depending on whether $g_d$ has zero, constant positive, or constant negative Ricci curvature. If $g_d$ is Ricci-flat, e.g. $g_d$ is the Minkwoski metric on $\mathbb{R}^d$, then the warp factor without the RS brane is $e^{A(z)} = L/z$, and the tension of the RS-brane has to be fine-tuned to the critical value
\be
\label{E:Tc}
	T_{RS} = T_c \equiv \frac{(d-1)}{4 \pi G_N L}\,,
\ee
where $L$ is the curvature radius of AdS$_{d+1}$. Since rescaling $z$ together with the coordinates on the slice is an isometry, the value of $z_*$  for the critically-tuned brane tension is undetermined. In this case, all values of positive $z_*$ are equivalent, and below we will usually work with $z_*=L$. For this choice of $z_*$, the induced metric on the RS brane is directly $-dt^2 + d\vec{x}^2$ without any additional constant scale factors.

In the case of the sub-critical tension $T_{RS}<T_c$ brane, the worldvolume metric $g_d$ of the brane has to be replaced with a negatively curved space \cite{Kaloper:1999sm,Kim:1999ja,DeWolfe:1999cp}. The slicing ansatz \eqref{slicing} together with the cut and paste procedure of \eqref{rs} can still be used to construct the full spacetime metric, which includes the backreaction of the brane. However, in order to satisfy the junction condition~\eqref{E:junction}, we now need $g_d$ to be a metric with constant negative Ricci curvature with radius $L$. For instance, $g_d$ can be the metric of pure AdS$_d$ with radius $L$. Before adding the RS-brane, Einstein's equations fix the warp factor to be
\be
\label{E:subWarp}
	e^A = \frac{1}{\cos(z/L)}\,.
 \ee
For super-critical tension $(T_{RS} >T_c)$, one needs $g_d$ to be a metric with constant positive Ricci curvature and radius $L$, e.g. the standard metric on dS$_d$. The warp factor is then given by $e^A=1/\sinh(z/L)$ before adding the RS-brane.

\begin{figure}[t]
\includegraphics[width=7cm]{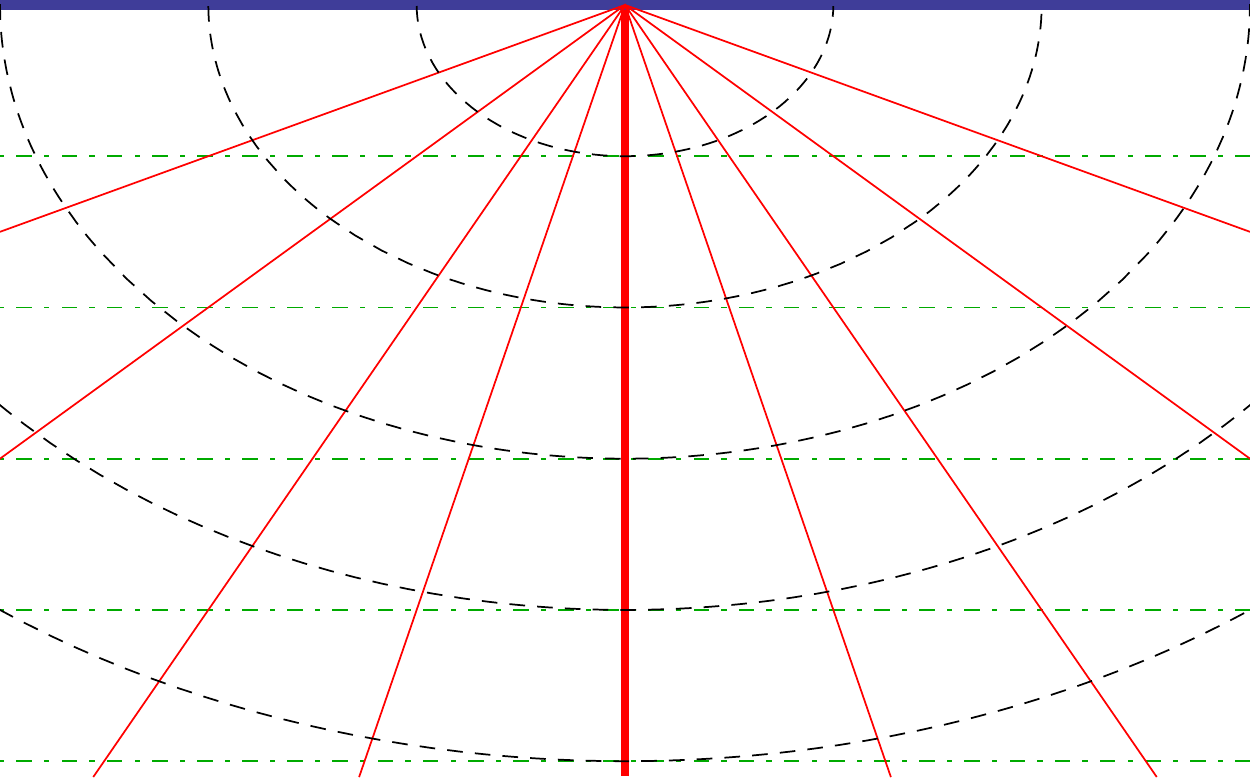}
\caption{ \label{fig:rsslicings} The Poincar\'e patch of AdS$_{d+1}$, in both flat-space and AdS$_d$ slicing. The green dash-dotted lines are the standard constant $z$ slices in flat slicing, where each slice is $d$-dimensional Minkowski space. The solid red slices are constant $z$ slices when the metric on the slice is taken to be AdS$_{d}$ with radius $L$. The dashed black lines correspond to fixed radial distance on the AdS$_d$ slice but arbitrary $z$ in the AdS$_d$ slicing. The solid blue line at the top of the plot is the conformal boundary of AdS$_{d+1}$. In flat slicing, it is naturally equipped with the $d$-dimensional Minkwoski metric. In AdS$_d$ slicing, it is the union of the two AdS$_d$ slices at $z/L=\pm \pi/2$, glued together along their common boundary. These boundary metrics are of course the same up to a Weyl transformation.}
\end{figure}

One qualitative difference between the sub-critical AdS$_d$ case and the others is that the warpfactor reaches a minimum value at $z=0$ and diverges at $z/L = \pm \pi/2$ corresponding to the AdS$_{d+1}$ boundary. We illustrate this in Figure~\ref{fig:rsslicings} by depicting both the standard flat slicing as well as the AdS$_d$ slicing of the Poincar\'e patch of AdS$_{d+1}$ (the corresponding picture for global AdS$_{d+1}$ can be found in e.g.~\cite{Karch:2000ct}). Without the brane, the boundary of AdS$_{d+1}$ in AdS$_d$ slicing is two copies of AdS$_d$ connected along their common boundary with one located at $z=\pi/2$ and the other at $z=-\pi/2$. Depending upon whether we take the slice to be the Poincar\'e patch or global AdS$_d$, these two joined copies of AdS$_d$ are conformally equivalent to $\mathbb{R}^d$ with the Minkowski metric, or $\mathbb{R}\times \mathbb{S}^d$. From this conformal map, we can learn that mSYM on these two copies of AdS$_4$ obey somewhat unusual boundary conditions.  As in the conformally equivalent case of two halves of Minkowski space, any excitation hitting the boundary of one AdS$_4$ gets completely transmitted into the other AdS$_4$, and so, we refer to these as transparent boundary conditions. For the subcritical case, $z_*$ is no longer a free parameter, but instead it measures the strength of the backreaction of the RS-brane. The backreaction is then fixed by the junction condition~\eqref{E:junction} in terms of the tension of the brane as
\be
\label{tensionrelation}
	\sin(z_*/L) = \frac{4 \pi G_N}{d-1} T_{RS} L\,.
\ee
Using this together with~\eqref{slicing},~\eqref{rs}, and~\eqref{E:Tc}, the induced metric on the RS-brane is AdS$_d$ with radius $l$ given by
\be
\label{E:inducedRadius}
	l^2 = \frac{T_c}{T_c-T_{RS}}L^2\,.
\ee
A similar relation also applies in the super-critical dS$_d$ case. According to \eqref{tensionrelation}, $z_* \rightarrow 0$ in the limit that the brane tension vanishes. In this case, the brane backreaction vanishes and the full spacetime goes back to undeformed AdS$_{d+1}$ in AdS$_d$ slicing. This is the ``probe limit" that can for example easily be realized by a single D5 brane probe \cite{Karch:2000gx} on AdS$_4\times\mathbb{S}^2$ inside of AdS$_5\times\mathbb{S}^5$.

One fact that will be important for us below is that the RS construction, \eqref{slicing} and~\eqref{rs}, provides a solution to Einstein's equations as well as the Israel junction condition for {\it any} Einstein metric $g_d$ with the appropriate warp factor. If $g_d$ is Ricci-flat, then we require the warp factor to take the form $e^A=L/z$; if $g_d$ has constant negative curvature with radius $L$, then $e^A=\sec(z/L)$; finally, if $g_d$ has constant positive curvature with radius $L$, then $e^A = 1/\sinh(z/L)$. As $g_d$ is naturally identified as the boundary metric, this gives an easy construction of holographic backgrounds realizing black holes on the brane. We will review this construction in more detail in the next Section.

The introduction of the critical RS-brane modifies the gravitational solution in three very important ways, all of which can readily be interpreted in the dual field theory (see \cite{Verlinde:1999fy,Gubser:1999vj} as well as the references therein). For one, part of the original AdS$_{d+1}$ space is removed. As indicated in \eqref{rs}, the Israel jump equations imply that for positive tension we remove the $z<z_*$ part of the AdS geometry.  That is from the field theory perspective, a UV cutoff has been introduced into mSYM given by $z^{-1}_*$. For our preferred choice of $z_*=L$, the cutoff in the field theory is $1/L$. Secondly, we doubled the geometry by gluing two cutoff slices to each other. This implies that the dual field theory is not a single cutoff mSYM based on a $SU(N)$ gauge group, but rather two copies of it or, equivalently, cutoff mSYM with gauge group $SU(N) \times SU(N)$. There is no problem with working with this doubled field theory, but if one prefers to have mSYM with a simple gauge group one can impose an orbifold projection in the bulk. By modding out by a $\mathbb{Z}_2$ action that identifies $z \rightarrow -z$, one therefore identifies the two cutoff AdS spaces. Most importantly, the RS setup adds a dynamical graviton to the field theory. From the bulk point of view, this can be seen by studying small fluctuations around the background. The discontinuity in the first derivative of the warpfactor gives rise to a delta function in the effective potential for the $d+1$ gravitational fluctuations that traps a massless $d$-dimensional graviton \cite{Randall:1999vf} in addition to the continuum of modes that represent the original $d+1$ dimensional graviton. In the holographic dictionary this is identified as an additional massless graviton coupled to the $d$ dimensional cutoff CFT. There is plenty of evidence for this picture to be correct. The Israel junction conditions are equivalent to $d$-dimensional Einstein's equations coupled to CFT matter~\cite{Emparan:1999wa}. If a black hole is placed in the $d+1$ dimensional bulk, the worldvolume of the RS brane is forced by the Israel equations to turn into an Friedmann-Robertson-Walker universe driven by an energy density that dilutes as a finite temperature CFT. When black holes are included on the cutoff slice with the RS brane, the bulk gravitational solution describes the field theory Hawking radiation backreacting on the $d$-dimensional metric \cite{Emparan:2002px} even though the properties of the Hawking stress tensor in this strongly coupled CFT are somewhat unconventional \cite{Fitzpatrick:2006cd}. In summary, the RS setup describes a duality between a cutoff CFT in $d$ dimensions coupled to a dynamical $d$-dimensional metric and a $d+1$-dimensional classical bulk theory containing $d+1$-dimensional gravity and the RS brane.

One special feature of the $d$-dimensional gravity in the RS setup is that its Newton constant $G_d$ can be viewed as being completely induced by quantum effects. From the bulk point of view, it can simply be calculated in terms of $G_N$ from the effective finite volume of the warped holographic dimension and the standard Kaluza-Klein formulas\footnote{The formulas we are quoting here are for the non-orbifold case, where the space contains two copies of cut-off AdS. If we only use a single copy, the volume of the holographic direction and hence $G_d$ is half of what is quoted here.} (using $z_*=L$):
\be
\label{gn}
	\frac{1}{G_d} = \frac{2 L}{d-2} \frac{1}{G_N}\,.
\ee
From the field theory point of view, we can understand this as a correction to the bare $1/G_d^{bare}=0$ due to a one-loop effect of the matter fields. The number of degrees of freedom in the field theory is set by $L^{d-1}/G_N$. The one-loop correction to $G_N$ is UV-divergent, but since our field theory is cut off at energy scale $1/L$, this translates into a contribution of $L^{2-d}$ per degree of freedom, giving (ignoring order one factors) \eqref{gn} as the correct induced $G_d$. This picture has been put forward early on in e.g. \cite{Hawking:2000da}. In theories in which gravity is completely induced, one expects that black hole entropy is entirely given by entanglement entropy across the horizon as both $G_d$ and the entanglement entropy arise from the same UV divergence. This has recently been confirmed to be the case for RS setups in \cite{Emparan:2006ni} using the Ryu-Takayanagi \cite{Ryu:2006bv} procedure to calculate the entanglement entropy of mSYM on a black hole background. The latter matching gives further support for the hypothesis that the RS model provides a holographic description for mSYM coupled to four-dimensional gravity.

Let us comment on the holographic interpretation for the case with a RS brane with tension below the critical value, $\it{i.e.}$ when the induced metric on the RS brane is AdS$_d$. Holography for this case was first discussed in \cite{Karch:2000ct,Karch:2000gx}, and the fluctuation analysis reveals that the trapped graviton in fact is massive. It can nevertheless be distinguished from the tower of massive KK modes (which are now discrete) in the near-critical limit where the radius of curvature on the brane~\eqref{E:inducedRadius} is much larger than the curvature radius $L$ of the bulk AdS$_{d+1}$. In this case, the trapped graviton has a mass that goes as $1/l^2$~\cite{Porrati:2001gx} as opposed to the $1/L^2$ scaling seen for the massive KK modes. In the probe limit, the trapped graviton just becomes indistinguishable from the massive modes. The holographic interpretation nevertheless goes through unchanged. Since the discussion involves two asymptotic regions, let us focus on the orbifolded case where we have just one copy of AdS$_{d+1}$ truncated at the RS brane. Note that the RS brane in this case only cuts out one of the two asymptotic regions at $z\to \pm( \pi/2-z_*)$. We started out with two copies of mSYM on two separate AdS$_d$ spaces linked via transparent boundary conditions. One of them remains untouched, while the other one gets replaced by cutoff mSYM coupled to mSYM. Unlike in Minkowski space, the propagator of a massive graviton in AdS$_d$ smoothly goes over to that of a massless graviton in the $m\rightarrow 0$ limit \cite{Porrati:2000cp}, which allows loop corrections to give massless gravitons a finite mass.  The analysis of \cite{Porrati:2001db} reveals that this will happen whenever the fields running in the loops have non-trivial boundary conditions. In particular, this is the case when mSYM is formulated on two copies of AdS$_4$ with transparent boundary conditions as discussed above.  Thus, this massive graviton is consistent with the interpretation of the RS geometry being dual to cutoff mSYM with a dynamical graviton. In fact, the preceding scenario provides a strong quantitative check of this proposal: the graviton mass can be computed, in mSYM, at weak coupling and one finds perfect agreement with the strong coupling analysis using the RS geometry \cite{Duff:2004wh}. While no precise non-renormalization theorem for this quantity has been proven, the fact that the mass agrees between strong and weak coupling is likely not an accident. It would be nice if one could prove that supersymmetry indeed protects the graviton mass in this scenario.

There is an alternative holographic interpretation in terms of a boundary conformal field theory (BCFT) describing the same sub-critical RS brane, as emphasized in \cite{Karch:2000ct,Karch:2000gx}. Since a single RS-brane only removes one of the boundary AdS$_d$ components, the full geometry can be thought of as being dual to the original CFT without a UV cutoff on the other AdS$_d$ component with non-trivial boundary conditions encoding the properties of the RS brane. This suggestion has recently been picked up again in terms of the AdS/BCFT correspondence \cite{Takayanagi:2011zk}. We will not make use of this holographic interpretation here, but we would like to emphasize that it is consistent with the point of view that RS is holography for field theories coupled to $d$-dimensional gravity. Since the RS brane contains a dynamical graviton on AdS$_d$, it should allow for a holographic interpretation. Acting with holography twice in this way naturally leads to the description of the system in terms of a BCFT. In this work we are expressly interested in studying quantum field theories coupled to quantum gravity, so we will not make this last step. Instead even when discussing the sub-critical tension branes, we will consider the RS setup as being dual to dynamical gravity coupled to a cut-off CFT.

\subsection{An EPR pair in mSYM coupled to gravity}
\label{S:EPRpair}

In our previous work \cite{Jensen:2013ora}, we constructed the holographic dual of a single quark-anti-quark ($q$-$\bar{q}$) pair in mSYM on flat Minkowski space in terms of an expanding semi-circular string, which was originally found in~\cite{Xiao:2008nr}\footnote{The solution in~\cite{Xiao:2008nr} can be obtained from the earlier construction in~\cite{Mikhailov:2003er} for the string worldsheet of an arbitrarily accelerating quark by gluing together two such solutions. This procedure has been used in~\cite{Caceres:2010rm,Garcia:2012gw} to construct an infinite family of solutions with similar worldsheet horizons.}. In terms of the standard Poincar\'e patch coordinates \eqref{slicing} with $e^A=L/z$, the embedding is given by
\be
	x^2 = t^2 + b^2 - z^2\,.
\ee
In the simplest case, the quark and anti-quark are test particles located at the boundary $z=0$ and are traveling along the trajectories $x(t)=\pm \sqrt{b^2+t^2}$. We see that at $t=0$ their separation is $2b$. The string worldsheet has worldvolume horizons at $z=b$, and thus, the induced geometry is that of a non-traversable wormhole. We concluded that in this setting the holographic dual of an EPR pair has a wormhole geometry. If one considers the endpoints of the strings to correspond to a dynamical quark and anti-quark, then the string ends on a flavor brane \cite{Karch:2003nh} located at $z_m$. In this case the string configuration is only consistent if the endpoints are pulled apart by an electric field $E=m/b$, where $m=\sqrt{\lambda}/(2 \pi z_m)$ is the quark mass and $\lambda = g_{YM}^2 N$ the `t Hooft coupling of the mSYM. As emphasized in \cite{Sonner:2013mba}, the most natural way to think of this pair is as being a $q$-$\bar{q}$ pair that was pair produced at $t=0$ by the same electric field that then accelerates them. In this interpretation, an instanton calculation determines $b$ in terms of $E$ and $m$ as the most likely initial separation of the pair.

\begin{figure}[h]
\includegraphics[width=7cm]{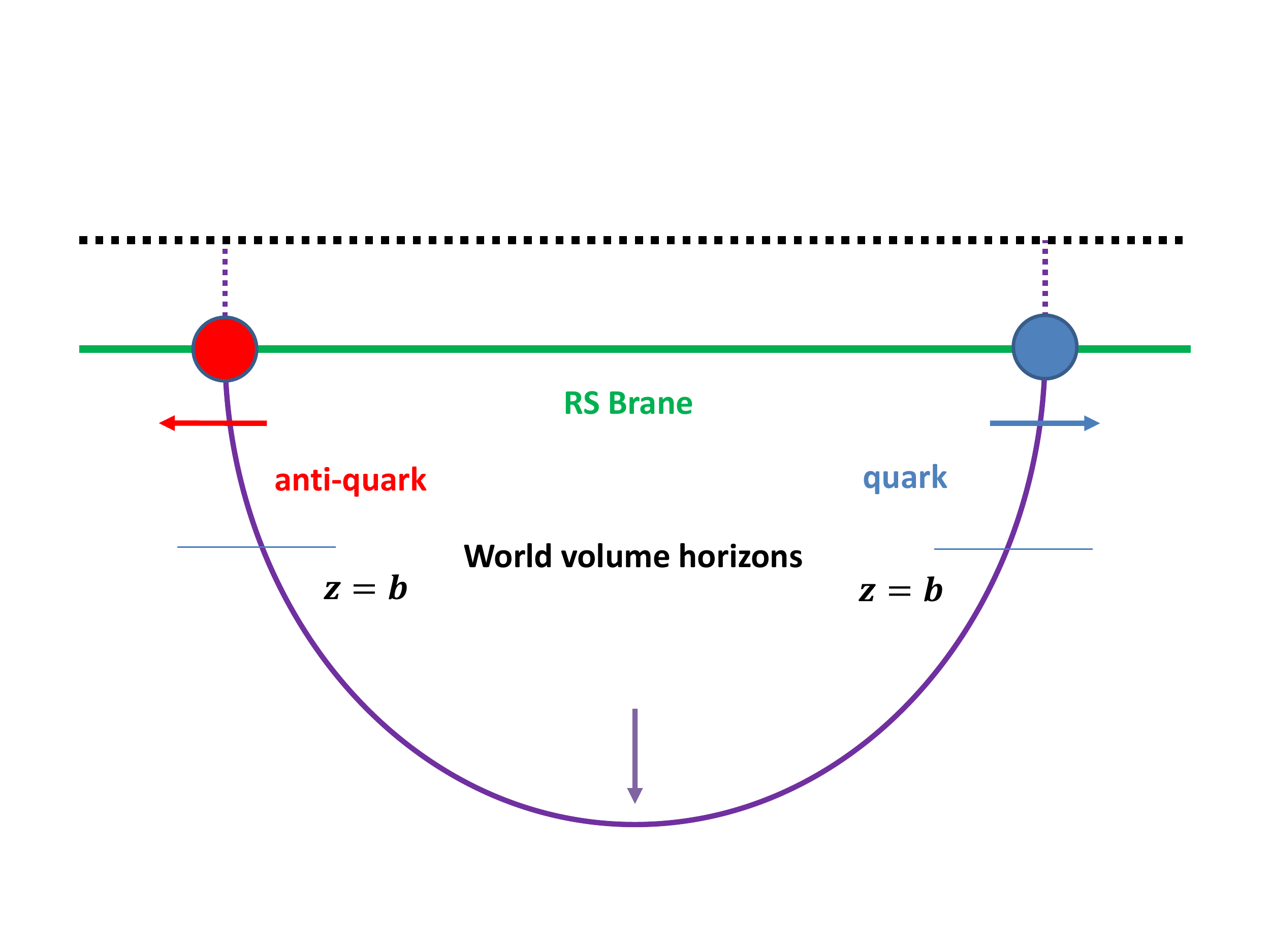}
\caption{String dual to a quark-antiquark pair in mSYM coupled to dynamical gravity, as realized by a critical RS brane. The string ends on the RS brane, and the endpoints are dual to a heavy dynamical quark and anti-quark, which uniformly accelerate away from each other for all time. When the initial separation $2b$ is large compared to the inverse cutoff scale $L$, there is a wormhole on the worldsheet of the dual string. However, the wormhole disappears when the initial separation is of order the inverse cutoff ($b<L$).}
\label{fig:dualpair}
\end{figure}

The same string is of course still a solution when we introduce the critical RS brane. The resulting holographic dual is displayed in Figure \ref{fig:dualpair}. Apparently even in the presence of dynamical gravity, the EPR=ER relation can be understood to imply that the holographic dual of an entangled pair is a wormhole. Obviously, this time the role of $z_m$ is played by $z_*=L$, as this is now the location where the string ends. In order to support the string profile, we therefore need an electric field on the RS brane with strength
\be
\label{E:RSelectric}
	E= \frac{\sqrt{\lambda}}{2 \pi} \frac{1}{L b} \,,
\ee
in units where the string endpoints have charge $\pm 1$. For this setup, we need an RS brane that supports a worldvolume gauge field just like a D-brane would, in addition to a tension. As the RS brane itself is an ad-hoc construct, there is no obstruction in introducing a worldvolume gauge field. However, the stress tensor associated with the RS worldvolume gauge field will now affect the Israel junction condition~\eqref{E:junction} so that our cut-off AdS background is no longer correct. We are now dealing with a theory that contains dynamical gravity, and so, in addition to leading to pair production of entangled $q$-$\bar{q}$ pairs, the electric field will also backreact on the metric leading to an expanding universe. In order to avoid this complication, we will work in the probe limit wherein the stress tensor of the worldvolume gauge field is asymptotically small compared to the tension in $L=1$ units. In this limit, the RS tension is $\mathcal{O}(1)$ in 5d Planck units, $\it{i.e.}$ $\mathcal{O}(N^2)$ in $L=1$ units with $N$ being the number of colors in the dual mSYM. However, we can choose that the coefficient of the $F^2$ term is negligible, $\it{i.e.}$ $\mathcal{O}(N)$ in $L=1$ units as it would be for a probe D-brane. As long as the string endpoint couples to the worldvolume gauge field with unit charge, the electric field~\eqref{E:RSelectric} is sufficient to create and accelerate the $q$-$\bar{q}$ pair.

One interesting aspect of the entangled pair in the RS setup is the following. For very large electric fields $E>\sqrt{\lambda}/(2\pi L^2)$, we lose the worldvolume horizon and so also the wormhole. Note that the parameter $b$ setting the initial quark separation\footnote{The true initial quark separation in the presence of the RS brane is no longer equal to $b$, but can be read off~\cite{Caceres:2010rm,Garcia:2012gw} by evaluating $x(z_m)=b^2-z_m^2$ at $t=0$.} is inversely proportional to the applied electric field.  So, the stronger the $E$-field, the smaller separation is needed between the particles to gain the equivalent of their rest energy in terms of electrostatic potential energy. We see that once $b<z_*=L$ the would-be worldvolume horizons move into the part of AdS$_{d+1}$ that is cut out by the RS brane and so our worldsheet wormhole disappears. This should not be surprising from the field theory point of view. Recall that $L$ is our short-distance cutoff. So we cannot resolve an entangled pair separated by a distance $b<L$ as being two particles. We can see, then, that the notion of local operators, which was required for the operational definition of entanglement in \eqref{connected}, is no longer appropriate when the operator separation is less than $L$. Note that according to~\eqref{gn} the cutoff length $L$ is much larger than the $4$d Planck length, so this is really a result of the cutoff and not of quantum gravity.

\section{Holographic duals of Hawking pairs}
\label{S:Hawking}

\subsection{Holography for mSYM on black hole backgrounds}

Having studied an entangled pair on flat Minkowski space in the presence of dynamical gravity, we wish to explore a scenario with an EPR pair in mSYM on a curved background with event horizons in the absence of dynamical gravity. Specifically, we will consider mSYM on a black hole background which, as we review below, does not Hawking radiate. Rather, the would-be Hawking radiation is confined. By studying an entangled pair with quark and anti-quark on opposite sides of the horizon, we will be describing a single entangled pair which would contribute to the Hawking radiation, a ``Hawking pair". As we will see, the radiation is confined in the sense that it takes some external force to keep the pair from falling into the horizon.

In the standard AdS/CFT dictionary the metric, $g_0$, on which mSYM is formulated is an input parameter. We demand that near the boundary the metric behaves as
\be
\label{fg} g= L^2 \frac{dz^2}{z^2} + \frac{L^2}{z^2} \left ( g_0 + \ldots \right )
\ee
where $\ldots$ stands for a power series in $z$ and $g_0$ is independent of $z$. For any given $g_0$, we can solve Einstein's equations subject to the requirement that the asymptotic structure of the solution take the form \eqref{fg}. We are interested in the case where $g_0$ describes a Schwarzschild or AdS-Schwarzschild black hole. One simple solution to Einstein's equations with these boundary conditions is the black string metric \cite{Chamblin:1999by} where one takes $d$-dimensional metric $g_d$ in~\eqref{slicing} to be that of a $d$-dimensional Schwarzschild black hole. As with this choice $g_d$ is still Ricci flat, the metric,~\eqref{slicing} still solves Einstein's equations. One problem with this setup is that it is subject to a Gregory-Laflamme instability \cite{Gregory:1993vy}. That is, thin black strings are unstable to spatial modulation, but since the warp factor $e^A=L/z$ for this solution becomes arbitrarily small, there is always some value of $z$ above which the string is unstable. This problem can be cured by taking $g_d$ to be the metric of an eternal AdS black hole, where the warpfactor is now
\be
\label{adswarp}
	e^A = \sec(z/L)\,,
\ee
and so takes a minimal value at $z=0$. It was shown numerically in \cite{Hirayama:2001bi} that for a sufficiently large AdS$_d$ black hole the corresponding black string is stable. One might wonder what is meant by sufficiently large. In \cite{Chamblin:2004vr} it was argued that if one adds a subcritical RS-brane to this setup with an AdS black hole on its worldvolume, then the onset point of the Gregory-Laflamme physics is associated with the Hawking-Page transition on the brane. The bulk Gregory-Laflamme instability sets in at exactly the temperature where, from the $d$-dimensional point of view, black holes become thermodynamically unstable, which is at a temperature slightly below the usual Hawking-Page phase transition in $d$ dimensions. In this section we focus on AdS$_d$ black branes, which are always above the Hawking-Page transition. As a result the corresponding $d+1$ dimensional black strings are always stable.

\begin{figure}[t]
\vspace{-.3cm}
\includegraphics[width=8cm]{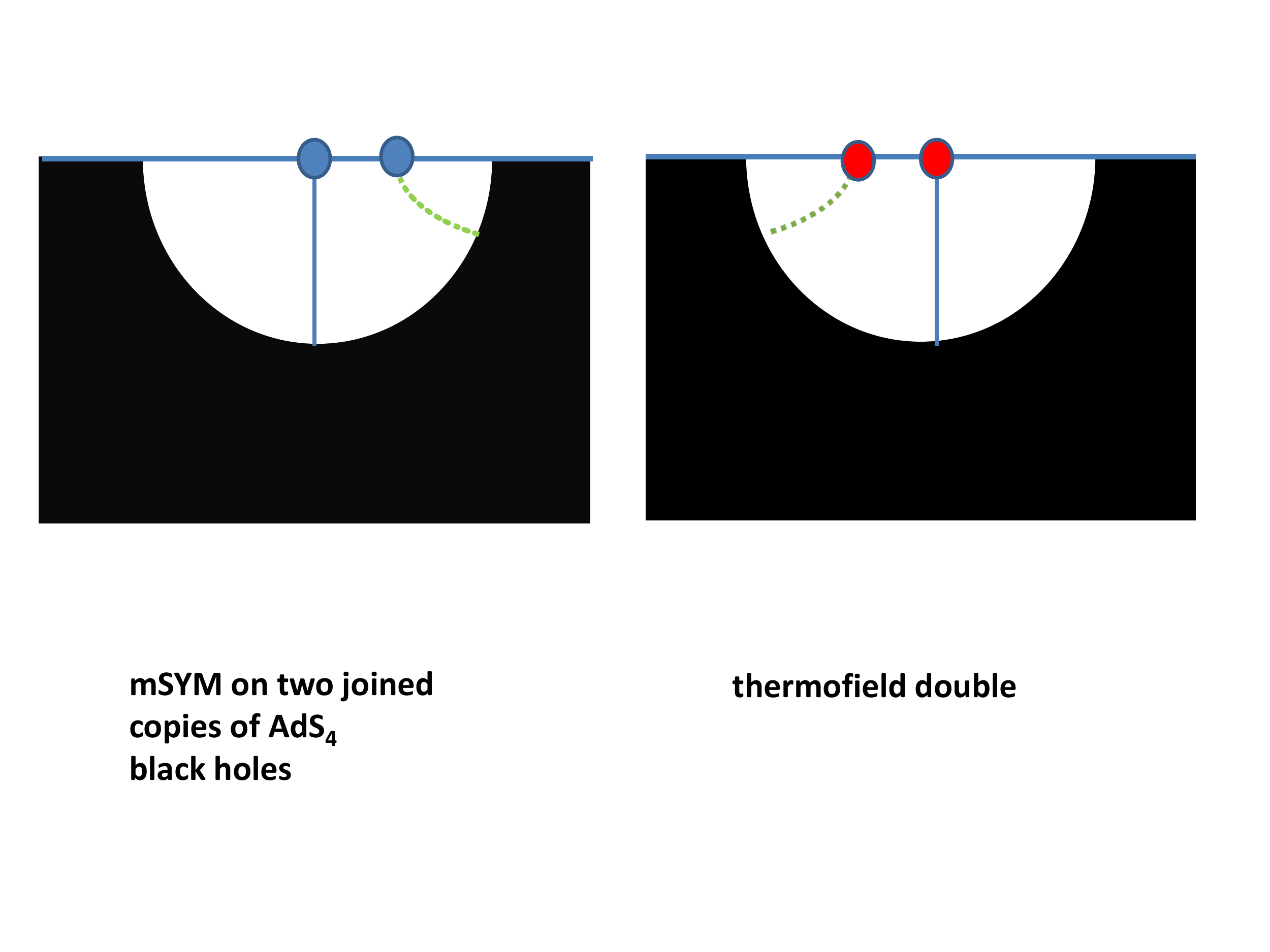}
\vspace{-.8cm}
\caption{\label{fig:adsstring} Strings dual to two different entangled Hawking pairs in the AdS black string geometry for two different values of $r_q=r_{\bar{q}}$. The circles are quark and anti-quark pair respectively, dashed and solid line are two possible strings connecting the two. The solid line is the analytic solution \eqref{straight} with $r_q=r_{\bar{q}}=\infty$. The figures show the $r$ and $z$ coordinates of \eqref{slicing} with $g_d$ given by the eternal AdS black brane \eqref{blackhole}. The time and transverse $\vec{x}$ directions are suppressed.}
\end{figure}

We depict the planar AdS black string in figure~\ref{fig:adsstring}. More concretely, the bulk metric is given by \eqref{slicing} with warpfactor \eqref{adswarp}. The metric on every slice, $g_d$, is that of an AdS$_d$ black brane,
\be
\label{blackhole}
	g_d = \frac{r^2}{L^2} \left (- f(r) dt^2 + d \vec{x}^2 \right ) + \frac{L^2}{f(r) r^2} dr^2\,,
\ee
where
\be
\label{black}
	f(r)=1 - \frac{r_h^{d-1}}{r^{d-1}}\,,
\ee
which has a horizon at $r=r_h$. The black string solution~\eqref{slicing} with this metric and warpfactor~\eqref{adswarp} has a Hawking temperature
\be
\label{E:bhT}
	T = \frac{(d-1)r_h}{4 \pi L^2}\,.
\ee
If we study the Euclidean theory, where $T$ sets the periodicity of the Euclidean time circle, we have to identify this Hawking temperature also as the temperature of the mSYM on this AdS black brane background.

\begin{figure}[t]
\includegraphics[width=7cm]{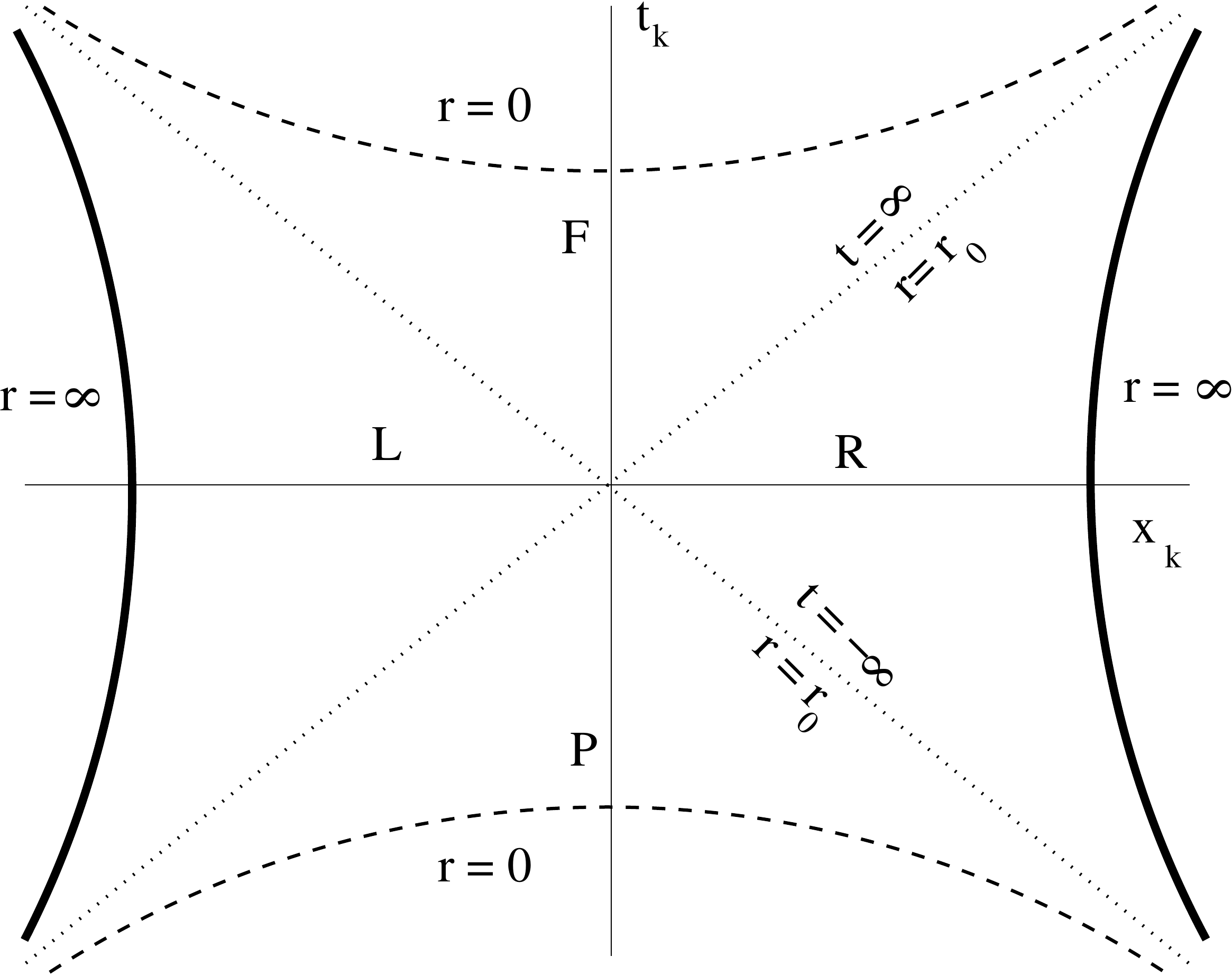}
\caption{\label{fig:kruskal} Figure from reference \cite{CasalderreySolana:2006rq} which depicts the causal structure of the Kruskal AdS$_d$ black hole present on each slice. Here, $r_0$ is the horizon $r_h$. The coordinate axes are the radial and time coordinate of the Kruskal coordinate system, which covers the entire space. The solid lines are the $r\rightarrow \infty$ boundaries of the extended black hole geometry. The dashed lines are the black hole singularities and the dotted lines are past and future horizons.}
\end{figure}

Since the spacetime described by~\eqref{blackhole} is not geodesically complete, we need to extend it. As usual, the coordinates in \eqref{blackhole} become singular at the horizon. Instead, one can use Kruskal coordinates to cover the extension as depicted in Figure \ref{fig:kruskal}. This extension has two horizons and two exterior regions connected by a non-traversable wormhole.  Note that we are implicitly putting mSYM on the entire Kruskal plane. The details of the change to Kruskal coordinates (see e.g. \cite{Fidkowski:2003nf,CasalderreySolana:2006rq}) are governed by the near-horizon behavior of the blackening factor
\be
	h(r) \equiv r^2 f(r) = h'_0 (r-r_h) + O(r-r_h)^2\,.
\ee
From \eqref{black} we have
\be
	h'_0 = (d-1) r_h\,.
\ee
To go to Kruskal coordinates, one first defines the `tortoise coordinate'
\be
	r_*= \int_0^r \frac{dr}{r^2 f(r)} + i \frac{\pi}{h'_0}\,,
\ee
where for $r>r_H$, we deform the integration contour slightly in the complex plane around the pole at $r_h$. The integration constant is chosen so that $r_*$ is real for $r>r_h$. The Kruskal coordinates are now defined as
\be
\label{coc}
	-e^{h'_0 r_*} = t_k^2 - x_k^2\,,
	\qquad
	\tanh \left ( \frac{h'_0}{2} t \right ) = \frac{t_k}{x_k}\,,
\ee
which brings the AdS black brane metric into the globally well-defined form
\be
\label{E:kruskal}
	g_d = \frac{1}{L^2}\left\{ G_k (- dt_k^2 + dx_k^2) + r^2 d \vec{x}^2\right\}\,,
\ee
with
\be
	G_k = \frac{4}{(h'_0)^2} h(r) e^{- h'_0 r_*(x_k,t_k)}\,.
\ee
The extended spacetime~\eqref{E:kruskal} can be equivalently represented by complexifying time in the original coordinates~\eqref{blackhole} on the R exterior~\cite{Fidkowski:2003nf}. The $R$ exterior is the $r>r_H$ part of~\eqref{blackhole}, while the F quadrant is the $r<r_H$ part, provided that we take the imaginary part of time $t$ in this quadrant to be $\text{Im}(t) = -\frac{\beta}{4}$ with $\beta = 1/T$. Similarly, the L exterior is described by the $r>r_H$ part of the metric~\eqref{blackhole} with $\text{Im}(t) = \frac{\beta}{2}$, and the P quadrant with $r<r_H$ and $\text{Im}(t) = \frac{\beta}{4}$.

The Kruskal extension of the metric \eqref{blackhole} has two asymptotic regions, each one being given by \eqref{blackhole}. If we were to use this black hole as the bulk geometry, we would identify the two regions as being dual to a CFT and its thermofield double. One should keep in mind that we currently use \eqref{blackhole} as the boundary geometry on which mSYM is formulated! As the metric \eqref{blackhole} appears on each slice, the full black string geometry also has two asymptotic regions. From the point of view of the observer living at the large $r$ boundary of \eqref{blackhole}, the second asymptotic region lies behind the horizon. The two asymptotic regions are connected by an ER bridge, a non-traversable wormhole, as depicted in Figure \ref{fig:adsstring}. By the same sort of arguments~\cite{Maldacena:2001kr} as those that lead to the conclusion that the eternal AdS black hole is dual to the thermofield double state\footnote{This interpretation has recently been called into question by \cite{Avery:2013bea}. The strongest argument in \cite{Avery:2013bea} seems to concern the late time behavior of correlation functions obtained from the eternal black hole background. It had already been noted in~\cite{Maldacena:2001kr} that at times parametrically large in $e^S$, where $S\sim N^2$ is the black hole entropy, the classical bulk picture breaks down: correlators obtained from the eternal black hole do not exhibit Poincar\'e recurrence as they ought. This is not a surprise, as the large $N$ and large time limits do not commute. But for any finite time there is substantial evidence that the eternal black hole geometry gives rise to correlators which obey all the theorems associated with thermal field theory, including the fluctuation-dissipation theorem. In here we assume that this standard interpretation of the eternal black hole is valid as proposed in~\cite{Maldacena:2001kr}.}, including the existence of a Euclidean instanton which smoothly patches onto the Lorentzian geometry at $t_k=0$, we conclude that IB strings on the extended geometry is dual to the Hartle-Hawking state of mSYM at temperature $T$ on the Kruskal extended black brane,
\be
|\Psi\rangle = \sum_i e^{-\beta E_i/2} |i\rangle_1\otimes |i\rangle_2\,,
\ee
where the sum runs over energy eigenstates. The two copies of mSYM live in the two exterior regions of the Kruskal extended black brane~\eqref{blackhole}.

\subsection{Holographic quark-antiquark pair}

\subsubsection{A simple holographic Hawking pair}

Now we would like to construct the holographic dual of an entangled pair in mSYM living on the planar $AdS_4$ black hole. We will concentrate on the case where the black brane horizon separates the quark ($q$) and anti-quark ($\bar{q}$), which we regard as a ``Hawking pair.'' For simplicity, we will consider the situation in which both the $q$ and $\bar{q}$ sit at the same position in the transverse $\vec{x}$ directions, which by translation invariance we take to be $\vec{x}=\vec{0}$. We denote the radial positions of the static $q$ and $\bar{q}$ as $r_q$ and $r_{\bar{q}}$ respectively. Note, that the quarks live on opposite sides of the ER bridge. For a given static $q$-$\bar{q}$ pair, there should be a unique fundamental string solution in the bulk black string geometry with one endpoint at $r_q$ and the other $r_{\bar{q}}$. This string is dual to the entangled Hawking pair. For the simple case $r_q=r_{\bar{q}}=\infty$, it is easy to write down an analytic solution for this string:
\be
\label{straight}
\vec{x}(r) =0\,, \quad \quad z(r) =0\,,
\ee
so the string is extended along $r$ and evolves in time. This string is depicted by the vertical blue solid line in Figure \ref{fig:adsstring}. It is easy to check that \eqref{straight} solves the equations of motion derived from the Nambu-Goto action in the black string metric with \eqref{blackhole} the metric $g_d$ on the slice. It was explicitly confirmed in \cite{CasalderreySolana:2006rq} that this string solution smoothly extends to the full Kruskal coordinates.

How does the string extend in the full Kruskal plane (see Figure~\ref{fig:kruskal})? In the same way that the extended geometry can be obtained by an analytic continuation of the geometry~\eqref{blackhole} as explained below~\eqref{E:kruskal}, so we can analytically continue the string using the coordinates~\eqref{blackhole}. The extended string extends across all four quadrants, all at $\vec{x}=\vec{0}$. The part of the string in the R quadrant is along $r>r_h$ and evolves in time $t$ at $\text{Im}(t)=0$. Similarly, the part of the string in the L quadrant is along $r>r_h$ at $\text{Im}(t)=-\beta/2$, and the parts in the P and F quadrants are along $r<r_h$ at $\text{Im}(t) = \beta/4, -\beta/4$ respectively. The evolution of the string, including its endpoints, can be described in each quadrant using the real part of the time $t$. For every time $t$ the explicit embedding $t_k(x_k)$ in the Kruskal plane can then be directly read off from the change of coordinates \eqref{coc}
\be
\label{sweep}
t_k = x_k \tanh \left ( \frac{h'_0}{2} t \right ) .
\ee
At $t=0$ the string lies along the horizontal $t_k=0$ axis and crosses from L to R at the bifurcation point where past and future horizon meet, $\it{i.e.}$ the origin of the Kruskal plane.  As time $t$ moves forward the right endpoint of the string moves up, whereas the left endpoint moves down in accordance with the natural notion of time for Kruskal plane.

Since the string embedding is invariant under the Killing vector that is equal to $\partial_t$ in the right exterior, the string and the rest of the right exterior are related by Wick rotation along $\partial_t$ to a smooth cigar geometry in which the string ends at the bottom where the geometry caps off. In this analytic continuation, we readily interpret the string as dual to a single external quark at $r_q\to \infty$ in the Euclidean thermal field theory.

As the string extends across the ER bridge, it directly inherits the geometry of an ER bridge on its worldvolume.  Thus, as in \cite{Jensen:2013ora}, we can conclude that the holographic dual of a Hawking pair has a wormhole!

\subsubsection{General quark position}

We would like to generalize the solution \eqref{straight} to more general $r_q$ and $r_{\bar q}$. For simplicity we want to focus on solutions that, like the simple example from \eqref{straight}, can be understood as the addition of a single external quark in the Euclidean thermal field theory. This only happens for symmetric configurations, $\it{i.e.}$ $r_q=r_{\bar q}$, so that the string is symmetric upon flipping $x_k\to - x_k$.

Rather than obtaining these solutions in the entire two-sided black string geometry, we will determine the part of the solution that lives in the R quadrant (and so the L quadrant by using the $x_k\to -x_k$ symmetry). Upon Wick rotation, the Euclidean on-shell action of the string $S_E$ gives the contribution of the $q$-$\bar{q}$ pair to the free energy $F$ of the thermal state, $S_E = F_{string}/T$. The extra free energy $F_{string}$ may be interpreted as the potential energy $F_{string}=V$ that the quark feels due to its color-neutralizing partner anti-quark behind the horizon.

Parameterizing the string embedding with either $z(r)$ or $r(z)$, the Nambu-Gotu action of the string is
\begin{align}
\begin{split}
\label{action}
	S_{NG} &=\frac{\sqrt{\lambda}}{2\pi L^2} \int  dr\,   \frac{1}{\cos^2 (z/L)} \sqrt{1 + f r^2 (z')^2}
	\\
	&=\frac{\sqrt{\lambda}}{2\pi L^2} \ \int dz \, \frac{1}{\cos^2 (z/L)} \sqrt{(r')^2 + f r^2} \,,
\end{split}
\end{align}
where $\lambda$ is the 't Hooft coupling of mSYM and the prefactor of $\sqrt{\lambda}/(2 \pi L^2)$ is the string tension $1/(2 \pi \alpha')$ after using $R^4=\lambda (\alpha')^2$. The string tension is large in the limit of large 't Hooft coupling, in which case we are justified in treating the worldsheet theory classically. From this point forward, we will work in units where $L=1$.

\begin{figure}[h]
\includegraphics[width=7cm]{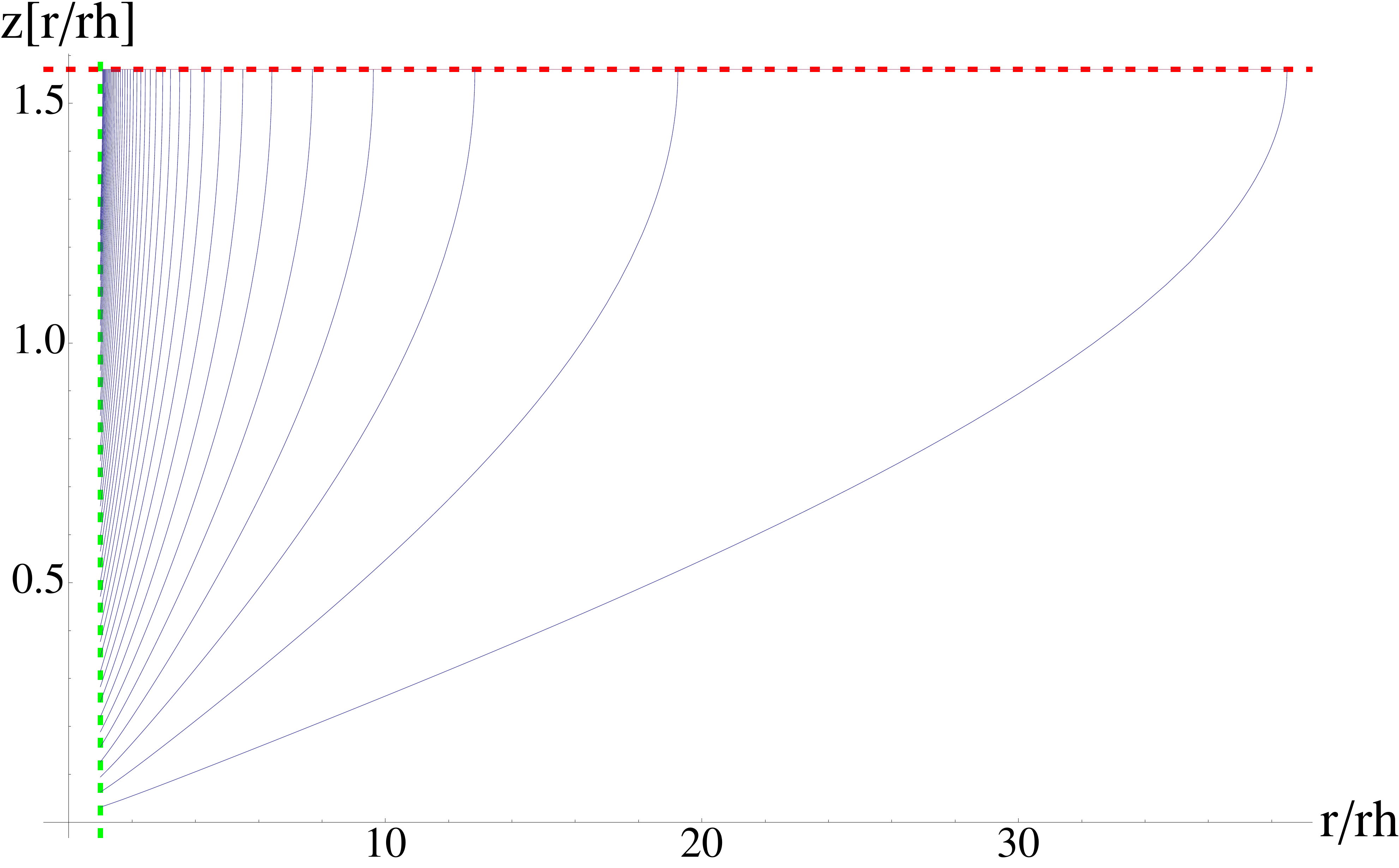}
\caption{\label{fig:StringConfig}The string configurations for a range of values of $r_q$ where the green, dashed vertical line is the location of the horizon $r_h =1$ and the red, dashed horizontal line represents the boundary $z= \pi/2$.}
\end{figure}

The Euler-Lagrange equation for the embedding $z(r)$ following from~\eqref{action} in general does not have analytic solutions, and so, we must employ numerical methods. In order to do so, we numerically integrate the equation of motion to determine $z(r)$, imposing regularity at the horizon $r=r_h$ and shooting up to the boundary. In the region near the horizon, the regular solution for $z(r)$ has a good Frobenius expansion in positive powers of $(r-r_h)/r_h$. There is also a logarithmically divergent solution which we discard by regularity. As usual, all higher coefficients in $(r-r_h)/r_h$ are uniquely fixed in terms of the horizon value of the embedding $z_*=z(r_h)$. In particular, $z'(r_h)$ is fixed in terms of $z_*$ as
\be
\label{regularity}
z'(r_h) = \frac{2}{h_0'} \tan(z_*)\,.
\ee
We scan through different values of $z_*$ and use the near-horizon Frobenius solution as initial conditions at a point $r_0$ near the horizon satisfying $r_0-r_h\ll r_h$, which we then numerically integrate up to $z=\pi/2-\delta z$ with $\delta z\ll 1$. We also solve the equation of motion near $\delta z=0$ in a Frobenius expansion, to which we match our numerical solution. From this matching we numerically determine the value of $r$ at which $z$ reaches $\pi/2$, which is $r_q$. A sequence of solutions with $r_q= \left[1.076,\,1.092,\,\ldots\,38.49\right]$ is displayed in Figure \ref{fig:StringConfig}.

\begin{figure}[h]
\includegraphics[width=7cm]{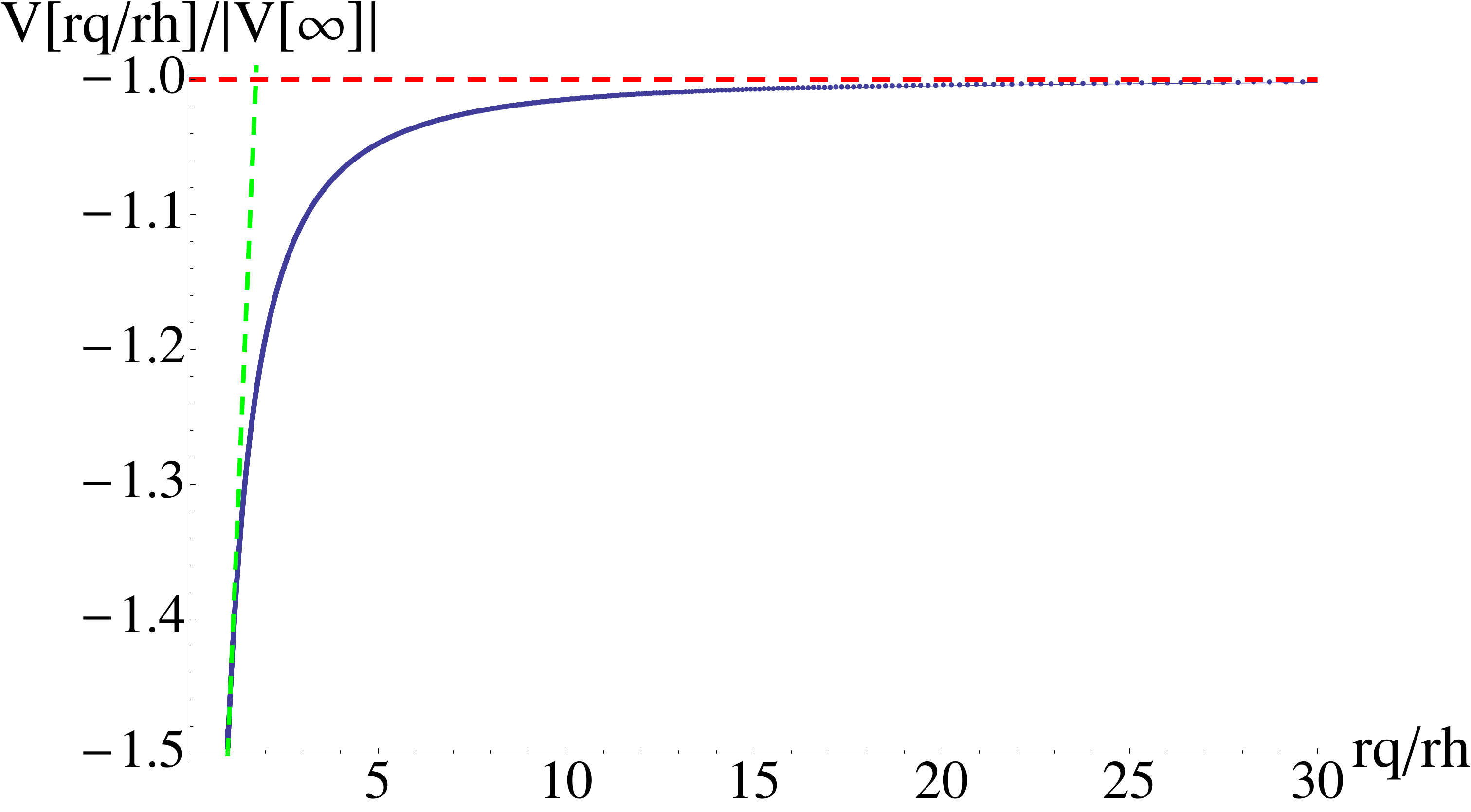}
\caption{The numerically obtained potential (blue, curved line), $V(r_q)$, plotted against the location of the quark $r_q$.  The dashed lines are the analytic solutions in the asymptotic regions $r_q \gg r_h$ (red, horizontal) and $r_q \sim r_h$ (green, vertical) given in the following section.}
\label{fig:Vrq}
\end{figure}

With the string embedding in hand, we can determine the corresponding free energy $F_{string}$. As always, the on-shell action diverges due to the infinite volume near the AdS-boundary, but this divergence can be regulated by holographic renormalization. We integrate up to a cutoff slice $z_{max}=\pi/2-y_{max}$ and add a volume counterterm on the near-boundary cutoff slice
\be
\label{counterterm}
S_{CT} = -\frac{\sqrt{\lambda}}{2\pi }\int dt \sqrt{-\gamma} =- \frac{\sqrt{\lambda}}{2\pi} \int dt \frac{r(z_{max})}{\cos(z_{max})}\,,
\ee
where $\gamma$ is the induced metric on the cutoff slice. The holographically renormalized action is
\be
S_{ren} = \lim_{y_{max}\to 0} (S_{NG}+S_{CT})\,.
\ee
Upon Wick rotating $t=-i t_E$, using $i S_{ren} = S_E$, and integrating over $t_E\in [0,1/T]$, we thereby obtain the contribution to the potential energy experienced by the quark due to the anti-quark behind the horizon
\be
V = T S_E\,.
\ee
We have computed the integral in $V$ numerically, and display our results in Figure~\ref{fig:Vrq}. We note that $V(r_q)$ displays the characteristics of a confining potential, in the sense that there is a steep linear barrier near the horizon and the potential always increases as $r_q$ increases. At large separation the potential linearly approaches the constant value $V\rightarrow -2\sqrt{\lambda}T/3$.

\subsubsection{Analytic expressions for small and large $r_q$}

For general $r_q$, we had to rely on numerical methods in solving a simple ODE in order to construct the holographic dual to a Hawking pair. We can construct approximate solutions analytically in the near-boundary ($r_q\gg r_h$) and near-horizon ($r_q\simeq r_h$) limits. These analytic solutions help to validate our numerics (see Figure \ref{fig:Vrq}) and facilitate interpretation of our results.

\vskip10pt

\noindent \underline{$r_q -r_h \ll r_h$:}
When $r_q$ is very close to the horizon, we can solve the string embedding perturbatively. To proceed we introduce a formal expansion parameter $\epsilon$ which we take to be $\epsilon \ll 1$. Rescaling the AdS ``radial coordinate'' $r$ as
\be
r = r_h + \epsilon \rho\,,
\ee
and expanding the embedding function $z(r)$ as
\be
\label{E:perturbativeZ}
z =- \frac{\pi}{2}+z_0(\rho) + \epsilon z_1(\rho) + \hdots\,,
\ee
we expand the Euler-Lagrange equation of motion for $z(r)$ following from~\eqref{action} in powers of $\epsilon$ around $\epsilon=0$. The $\mathcal{O}(\epsilon^0)$ part of the equation is a first-order nonlinear equation for $z_0$,
\be
\label{E:RHleading}
z_0' = - \frac{2}{3r_h}\cot(z_0)\,,
\ee
and the higher $\mathcal{O}(\epsilon^n)$ terms are first-order linear equations for $z_n$, wherein the lower-order solutions $z_{i<n}$ appear as source terms. Imposing regularity at the horizon $\rho=0$, we find for instance
\begin{align}
\label{E:RHsol}
z_0 & = \text{arccos}\left[ \exp\left( \frac{2(\rho-(r_q-r_h))}{3r_h}\right)\right]\,,
\\
\nonumber
z_1 & = \frac{2}{9}\frac{\rho^2-(r_q-r_h)^2}{r_h^2}\left[\exp\left( \frac{4(r_q-r_h-\rho)}{3r_h}\right)-1\right]^{-\frac{1}{2}}\,,
\end{align}
where we have labelled the integration constant $r_q$ so that the $z_i$ are valued on the domain $\rho \in [0,r_q-r_h]$. The first-order correction parametrically goes as $(r_q-r_h)/r_h$ times the leading order solution $z_0$.

To determine the potential energy corresponding to this solution, we evaluate the Nambu-Gotu action~\eqref{action} on the perturbative solution~\eqref{E:perturbativeZ} and substitute $\varepsilon=1$. We then expand the action in inverse powers of $r_h$ around $r_h=\infty$, keeping $\rho$ and $r_q-r_h$ fixed. The leading-order and NLO terms~\eqref{E:RHsol} determine the string action to $\mathcal{O}(r_h^{0})$. We then integrate over $\rho$ up to a cutoff $\rho_c = r_q-r_h-\delta$, add the appropriate counterterm~\eqref{counterterm} on the slice, and take the $\delta \to 0$ limit to get a renormalized action
\be
S_{ren} = \frac{\sqrt{\lambda}}{2\pi}\int dt \, r_h\left[ -\frac{3}{2} + \frac{2}{3}\frac{r_q-r_h}{r_h}+ \mathcal{O}\left( \frac{(r_q-r_h)^2}{r_h^2}\right)\right]\,.
\ee
Wick-rotating time and integrating over $t_E \in [0,1/T]$ with $T$ given by~\eqref{E:bhT}, we find
\be
S_E = \sqrt{\lambda}\left\{-1 + \frac{4}{9}\frac{r_q-r_h}{r_h}+\mathcal{O}\left( \frac{(r_q-r_h)^2}{r_h^2}\right)\right\}\,.
\ee

\vskip10pt

\noindent \underline{$r_q \gg r_h$:} For large separation between quark and horizon, the string will approximately follow an $r_h=0$ solution, and we can treat the effects of $r_h$ as a small perturbation, at least for $r\gg r_h$. That is, the endpoint of a string near the spatial boundary ($r\rightarrow\infty$) should not be sensitive to the presence of the horizon.  For $r_h=0$, the metric on the slice is simply AdS$_4$ and the full spacetime \eqref{slicing} with warpfactor \eqref{adswarp} is the Poincar\'e patch of AdS$_{5}$ in AdS$_d$ slices. The latter can equivalently be written in standard Poincar\'e patch coordinates:
\be
	\label{poincare}
	g = \frac{du^2 + dw^2 + g_{2,1} }{u^2}\,,
\ee
where $g_{2,1}$ is the Minkowski metric on $\mathbb{R}^{2,1}$. In these coordinates, it is very easy to see that a string straight hanging down at a fixed position in $w$ and sitting at an arbitrary spatial point in $\mathbb{R}^{2,1}$,
\be
w(u) = r_q^{-1}\, , \quad \vec{x}=0\,,
\ee
solves the string equations of motion. The change of coordinates that takes us from \eqref{poincare} to our black string metric \eqref{slicing} in terms of $z$ and $r$ is given by
\be
u = \frac{\cos(z)}{r}\,, \quad w = \frac{\sin(z)}{r}\,.
\ee
From this, we can easily read off the leading $r_h=0$ solution for the static string as
\be
\label{E:largerqSol}
r = r_q \sin(z)\,.
\ee
At $r_h$ nonzero but small compared to $r_q$, the $r_h=0$ solution~\eqref{E:largerqSol} gets $\mathcal{O}(r_h^3)$ corrections for $r\gg r_h$, or equivalently $z\gg r_h/r_q$, and $\mathcal{O}(r_h)$ corrections near the horizon with $r\sim r_h$, or $z\sim r_h/r_q$. Accounting for the fact that the integral over Euclidean time introduces a factor $\sim 1/r_h$, these corrections to the string solution lead to corrections to the on-shell action~\eqref{action} which go as $\mathcal{O}(r_h^2)$ arising from the $z\gg r_h/r_q$ region of integration, and as $\mathcal{O}(r_h)$ arising from the near-horizon region. So to $\mathcal{O}(r_h^0)$, the on-shell action is that of the $r_h=0$ embedding~\eqref{E:largerqSol} evaluated on the domain $z\in [\arcsin(r_h/r_q)\sim r_h/r_q,\pi/2]$. It is straightforward to evaluate the resulting action, which we find to be
\be
S_{E} = - \frac{2\sqrt{\lambda}}{3} + \mathcal{O}\left( \frac{r_h}{r_q}\right)\,,
\ee
which comes entirely from the lower bound of integration in~\eqref{action}.

\section{Asymptotically flat black holes}
\label{S:asyFlat}

In Section~\ref{S:Hawking}, we studied mSYM and Hawking pairs on two-sided AdS-black brane backgrounds. In principle, we can also put mSYM on an asymptotically flat black hole background, such as the two-sided Schwarzschild black hole, at a temperature $T$ equal to that of the black hole. What is the corresponding gravity dual of this equilibrium state? As we mentioned at the beginning of section~\ref{S:Hawking}, one possible answer is that the dual geometry is given by the metric~\eqref{slicing} with the Schwarzschild metric on the constant-$z$ slices. This bulk solution describes a black string, where the radius of the string shrinks as $1/z$ away from the AdS boundary. Unsurprisingly, this string suffers a Gregory-Laflamme instability~\cite{Gregory:1993vy}, which indicates that the dual to mSYM on Schwarzschild is something else.

Whatever the true dual geometry is, it must exhibit three crucial features. (See e.g.~\cite{Hubeny:2009ru} and the review~\cite{Marolf:2013ioa}.) First, the field theory on Schwarzschild is in thermal equilibrium, and so the gravity dual must be a time-independent geometry with a regular Euclideanization. Second, far away from the Schwarzschild horizon, the boundary geometry is flat and the mSYM is at nonzero temperature. So the dual geometry is that of an AdS black brane at temperature $T$ far from the Schwarzschild horizon. Third, the horizon on the boundary must be extended into the bulk. There are then two possibilities for the bulk geometry. The first is a so-called ``black funnel,'' with a single bulk horizon which continuously connects the Schwarzschild horizon on the boundary to the black brane horizon far away. Alternatively, the bulk horizon might be disconnected into two pieces, one which hangs from the AdS boundary above a (deformed) planar horizon. The latter is known as a ``black droplet.'' See Figure 3 of~\cite{Marolf:2013ioa} for a cartoon of the funnels and droplets. While black funnels have been numerically constructed in~\cite{Santos:2012he}, the same cannot be said for black droplets (at least, at a temperature $T$ equal to that of the deformed black brane). As a result it is not known which solution has the smallest on-shell Euclidean action, and therefore is dual to mSYM on Schwarzschild.

Without the gravity dual in hand, we cannot numerically study strings dual to Hawking pairs. Nevertheless, we can use knowledge of the dynamics of strings on AdS black branes (as in~\cite{Herzog:2006gh,Gubser:2006bz} for heavy quarks and~\cite{Gubser:2008as,Chesler:2008uy} for light quarks) to draw some qualitative conclusions about entanglement and wormholes in this setting.

To proceed, we consider a related problem in the form of entangled heavy $q$-$\bar{q}$ pairs in mSYM at nonzero temperature. We take the quarks to have a thermal mass $m$. In the gravity dual, we consider strings in the AdS$_5$ black brane~\eqref{blackhole}, where the endpoints hang from a flavor brane~\cite{Karch:2003nh} at $r=r_m$, which is related to the thermal mass as $m = \sqrt{\lambda}(r_m-r_h)/(2\pi)$~\cite{Herzog:2006gh}. This setup shares many qualitative features with $q$-$\bar{q}$ pairs in the dual of mSYM on Schwarzschild, insofar as both the black droplets and funnels asymptote to black branes far away from the Schwarzschild radius. Now, as in the $T=0$ entangled $q$-$\bar{q}$ pair we reviewed in Subsection~\ref{S:EPRpair}, turn on an electric field $E$ on the flavor brane. This has the effect of creating entangled $q$-$\bar{q}$ pairs via Schwinger pair production, after which the quarks are accelerated away from each other. From a combination of numerical simulations and analytical results, we know that the quarks are initially accelerated, but at late times settle down to constant-velocity trajectories as the electric force on the quarks is balanced by what in mSYM should be understood as a thermal drag force~\cite{Herzog:2006gh,Gubser:2006bz}. In this late-time asymptotic regime, the part of the string worldsheet near the endpoints can be obtained analytically; this is the so-called ``trailing string.'' The late-time velocity $v$ may be computed from this worldsheet and is related to the electric field and temperature as
\begin{equation*}
E = \frac{\pi \sqrt{\lambda}T^2}{2}\frac{v}{\sqrt{1-v^2}}\,.
\end{equation*}

Crucially for us, the worldsheet has a two-sided horizon which is formed during its evolution; at late times it is located at $r_* = r_h(1-v^2)^{-1/4}$~\cite{Gubser:2006nz}. The physical interpretation of these and other worldvolume horizons in AdS/CFT has been unclear\footnote{In~\cite{Chernicoff:2013iga} it has been argued that worldsheet horizons appear whenever the quark system emits radiation to infinity. For our $q$-$\bar{q}$ pair, this implies that the pair is effectively unbound, with each of the partners emitting gluon radiation. This picture seems to be consistent with our findings. We interpret the fact that the string remains connected while the pair unbinds as a reflection the non-trivial entanglement in the state.}. Here, following~\cite{Maldacena:2013xja,Jensen:2013ora}, we conjecture that the worldsheet horizons encode the color entanglement between the quark and anti-quark. The obvious difference between this example and the $T=0$ accelerating pair we reviewed in Subsection~\ref{S:EPRpair} is that, while the uniformly accelerating $q$-$\bar{q}$ were entangled and out of causal contact for all time, the entangled pair at $T\neq 0$ is never out of causal contact and so, by the definition we gave in the Introduction, is not an example of EPR entanglement. However, at nonzero temperature, thermal screening typically renders spacelike separated correlators to be exponentially suppressed at distances longer than the correlation length. As a result, perhaps the right definition of EPR entanglement at nonzero temperature is that spacelike separated correlations do not die off exponentially beyond the correlation length. In this instance, we do not know how the quark and anti-quark are correlated at late times. Although, the fact that the string smoothly connects them allows the possibility that they are correlated even arbitrarily far apart.

Now consider an entangled $q$-$\bar{q}$ pair in the dual to mSYM on the Kruskal extension of the Schwarzschild black hole. Let us study a static configuration with the quark in the right Schwarzschild exterior and the anti-quark in the left exterior region. Then, just as in the Hawking pair we studied in Section~\ref{S:Hawking}, there is a two-sided wormhole on the worldsheet of the dual string where the string crosses the bulk horizon. Now suppose that, as above, the quarks are heavy so that the dual string ends on a flavor brane, and moreover put an electric field on the brane which pulls the quark away from the Schwarzschild horizon toward asymptotic infinity. When the quark is far away from the horizon, then the geometry is approximately that of the AdS$_5$ black brane, and the string worldsheet near the quark is that of the ``trailing string'' we mentioned above. During the evolution of the $q$-$\bar{q}$ pair, the worldsheet horizon closest to the quark moves out and away from the bulk horizon, and ultimately sits at $r_*$, moving toward asymptotic infinity at the same velocity $v$ as the quark. As with the $q$-$\bar{q}$ pair in mSYM at $T>0$, we expect that the worldsheet horizon here encodes the entanglement between the quarks. In both this setup and in the $T>0$ configuration, a contribution of the string to the mutual information of regions surrounding the quark and anti-quark that does not die off exponentially with distance would quantify their long-distance entanglement. We leave these and related inquiries to future work.

\section{Hawking pairs in RS setups}
\label{S:HawkRS}

In the previous Sections, we have first shown how to construct the holographic dual of an entangled pair in mSYM on a flat space with dynamical gravity and second on a curved black hole background in the absence of dynamical gravity. It is natural to bring the two constructions together to describe entangled Hawking pairs in a theory with dynamical gravity. Given our preceding results, this is almost trivial, at least for the case where we put mSYM on the AdS black brane background~\eqref{blackhole}, which we focus on for the entirety of this Section. We can simply add a RS brane at $z=z_*$ into the black string geometry. Our string solutions of Section~\ref{S:Hawking} are unaffected by this insertion. In order for the string to consistently end on the RS brane, we once more need to turn on worldvolume electric fields. Provided that the electric field is a probe on the RS brane, this can again be done without effecting the string solution by turning on appropriate worldvolume gauge fields as in Section \ref{S:RS} . The holographic dual of a Hawking pair in mSYM coupled to dynamical gravity in the presence of a black hole is a horizon crossing string in the bulk black-string solution. It inherits the structure of an ER bridge on its worldvolume.

There are some interesting lessons to be learned here about black holes in AdS-sliced RS models. One puzzling aspect of the stable AdS-sliced black string solution that was emphasized in \cite{Gregory:2008br} is that the metric on the RS slice is the standard black hole metric, completely unmodified. One should have expected that in a theory with dynamical gravity, the backreaction of the Hawking radiation on the black hole geometry should modify it. In fact, it was shown in \cite{Gregory:2008br} that the stress tensor for mSYM on the AdS-Schwarzschild background \eqref{blackhole} is proportional to the background metric. Thus, it only gives a renormalization of the curvature radius on the brane, which is not what one would expect from radiation. So, it is consistent that there is an absence of non-trivial backreaction as there is no Hawking radiation. In sharp contrast, at weak coupling one does instead find the expected Hawking radiation \cite{Gregory:2008br}.

How can this be explained? One potential scenario was pointed out in \cite{Fitzpatrick:2006cd}, which focused on the analogous problem for the flat-sliced black string. While this black string geometry is unstable as discussed above, it still represents a valid solution to bulk gravity with the right boundary data, and so, the model does constitute a valid dual for an (unstable) state of mSYM on a standard Schwarzschild black hole background. The claim of \cite{Fitzpatrick:2006cd} is that the effective number of degrees of freedom accessible to the Hawking radiation is not $\CO(N^2)$ but $\CO(1)$. This implies that the thermal Hawking radiation is still emitted at the standard Hawking temperature with an $\CO(1)$ constant of proportionality in the resulting Stefan-Boltzmann law. This is consistent with the calculation of \cite{Gregory:2008br}. The classical geometry only captures the leading $\CO(N^2)$ piece of the boundary stress tensor. The $\CO(1)$ terms would map to bulk quantum effects.  In order to calculate the stress tensor of Hawking radiation on the boundary, one needs to calculate the Hawking radiation of the black string metric in the bulk.

How exactly this ``confinement" mechanism works was not quite clear. This issue was recently revisited in \cite{Hubeny:2009rc} where it was shown that the related question in the double-Wick rotated geometry with an AdS-Soliton (\eqref{blackhole} on the slice with the roles of $t$ and one of the $\vec{x}$ exchanged) has a straightforward answer. In this model, mSYM is driven into a confining phase by the non-supersymmetric circle compactification. Since the geometry can be analytically connected back to the black string geometry, the analysis of \cite{Hubeny:2009rc} indeed postdicts the fact that the black hole should also effectively act as if it were confining, but again, how this happens dynamically is not clear. Obviously far away from the black hole, mSYM is in its standard deconfined phase. What is required to avoid $\CO(N^2)$ energy densities in the Hawking radiation is that colored Hawking pairs are unable to be emitted from the horizon - they should see an effective flux tube pulling them back in. Only color neutral particles can be emitted in the Hawking radiation.

However, we have the answer to these concerns at hand.  Let us recall the results from the previous sections. In studying the holographic dual to a heavy $q\bar{q}$ pair ($M_q\sim\sqrt{\lambda}$) separated by the horizon of a black hole, we have found that potential, which we plotted in Figure \ref{fig:Vrq}. It is confining and is entirely due to the quark interaction with the color-charged horizon. In this way, we see that the AdS black string does not Hawking radiate any colored particles according to an observer at infinity.

\section{Cosmological Horizons}
\label{S:cosmo}

In the previous sections we studied mSYM on a black brane background, with and without dynamical gravity. We saw that an entangled Hawking pair made of a colored quark and anti-quark on opposite sides of a horizon has a dual description in terms of a string worldsheet with an ER bridge on its worldvolume. One would expect a similar construction to also work in the case of cosmological horizons. In this section, we will show that this is indeed the case for the simplest cosmological scenario we can study with our setup: mSYM on dS$_4$. For our consideration, we will focus on the case without dynamical gravity, but as in the previous sections inclusion of dynamical gravity via an RS brane is straightforward and does not alter the solution.

\begin{figure}[h]
\includegraphics[width=7cm]{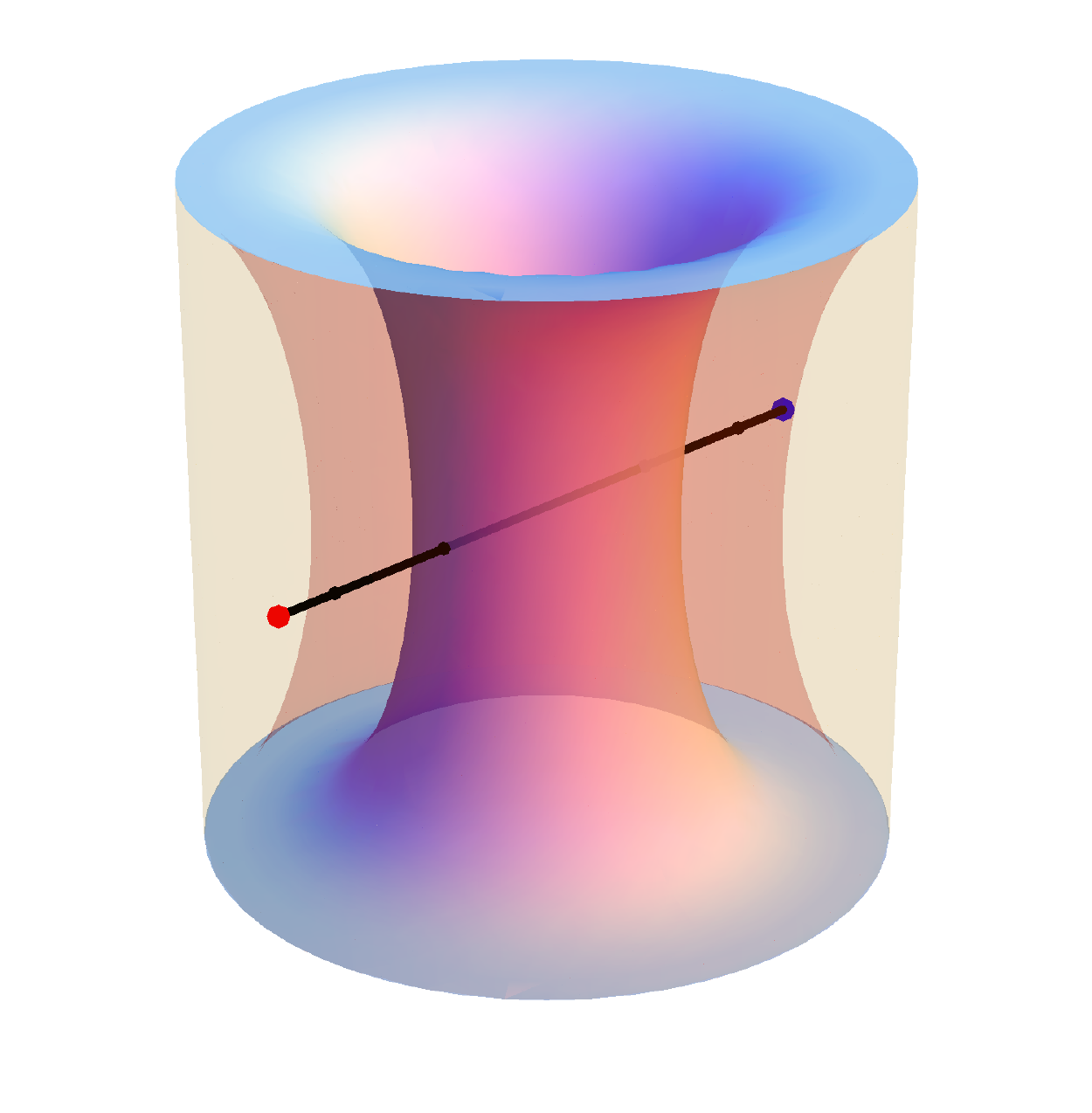}
\caption{Global AdS space with two global dS slices displayed. The black line is a string piercing global AdS along a constant angle, connecting the north and south pole of the boundary sphere.}
\label{fig:ds}
\end{figure}

Fortunately for us, the dual to mSYM on dS$_4$ is just IIB string theory on global AdS$_5\times\mathbb{S}^5$, written in de Sitter slicing. To study mSYM on dS$_4$ we once more employ the metric ansatz \eqref{slicing}, now with dS$_d$ on the slice by requiring
\be
	\label{globalds1}
	e^A = \frac{1}{\sinh(z/L)}\,.
\ee
To see what part of AdS$_{d+1}$ this coordinate system covers, we need to commit to a form of the dS$_d$ metric on the slice. If we chose global coordinates,
\be
	\label{globalds2}
	g_d = -d \tau^2 + L^2 \cosh^2(\tau/L) g_{\mathbb{S}^{d-1}}\,,
\ee
for $g_{\mathbb{S}^{d-1}}$ the metric on a unit-radius $d-1$-sphere, this slicing corresponds to a sequence of ``hyperboloids" covering a region of global AdS$_{d+1}$ very similar to the L and R wedges of the Kruskal plane in the black hole metric.\footnote{The remaining part of AdS is covered by an FRW universe with hyperbolic slices. The strings we are interested in this work do not reach into this region, so we will not discuss it in any more detail.} The slicing ends at horizons at $z =\infty$, and two of these global dS$_d$ slices are displayed in Figure \ref{fig:ds}. Were we to include an RS brane in this slicing, we would need to have a supercritical tension brane. The position of the brane will be determined by
\be
\cosh(z_*/L) = \frac{4 \pi G_N}{d-1} T_{RS} L\,.
\ee

For a local observer, dS$_d$ space appears to have a cosmological horizon. This can be seen using static coordinates for a patch of dS$_d$,
\be
	\label{staticds}
	g_d = - f(r) d t^2 + \frac{dr^2}{f(r)} + r^2 g_{\mathbb{S}^{d-2}}\,,
\ee
where the metric function takes the form
\be
	f(r) = 1 - \frac{r^2}{L^2}\,.
\ee
This space has a horizon at $r=L$ which appears to the static observer to be at the temperature $T=1/(2 \pi L)$. Note that this temperature is an observer dependent effect, and so, unlike the case of Hawking radiation from a black hole, we do not expect it to backreact on the metric. Nevertheless, we can study the analogue of a Hawking pair in this case as well with a quark and an anti-quark on opposite sides of the horizon.

To construct such string solutions, we obtain them first in global AdS$_{d+1}$ and then map the solutions to strings in the dS$_d$ slicing. The metric of global AdS$_{d+1}$ is
\be
	\label{globalads}
	g= - F(\tilde{r}) d \tilde{t}^2 + \frac{d \tilde{r}^2}{F(\tilde{r})} + \tilde{r}^2 g_{\mathbb{S}^{d-1}}\,,
\ee
where
\be
	F(\tilde{r}) = 1 + \frac{\tilde{r}^2}{L^2}\,,
\ee
and here, as well as in \eqref{globalds2}, we pick a representative for the $d-1$-sphere metric
\be
	g_{\mathbb{S}^{d-1}}= d \theta^2 + \sin^2(\theta) g_{\mathbb{S}^{d-2}}\,.
\ee
We consider static string embeddings extending along $\tilde{t}$ and sitting at a point in the $\mathbb{S}^{d-2}$. Its embedding is parameterized by $\theta(r)$, which connects the $q$ and $\bar{q}$ at two different points on the $d-1$ sphere. The Nambu-Gotu action for such an embedding is given by
\be
	S = \frac{1}{2\pi \alpha'}\int dt dr \sqrt{1+ F \tilde{r}^2 (\theta')^2}\,,
\ee
and the Euler-Lagrange equation for $\theta(r)$ is solved by the first integral
\be
	\label{solution}
	\theta' = \frac{\alpha}{ F \tilde{r}^2 \sqrt{1 - \frac{\alpha^2}{F \tilde{r}^2}}}\,.
\ee
Here $\alpha$ is an integration constant setting the separation in $\theta$ between the two endpoints. The simplest solution is given by $\alpha=0$, which describes a string going straight from the north pole to the south pole of the boundary sphere of global AdS$_{d+1}$. This solution is displayed in Figure \ref{fig:ds}. Clearly for every pair of $q$ and $\bar{q}$ positions, $\theta_q$ and $\theta_{\bar{q}}$, there is exactly one solution of the form \eqref{solution}. These solutions can easily be rewritten in terms of the global dS slicing of AdS$_{d+1}$,~\eqref{slicing} together with \eqref{globalds1} and \eqref{globalds2}, by noting that the two coordinate systems have the same $g_{\mathbb{S}^{d-2}}$ factor So, $\theta$ is the same in both coordinate systems. The change of coordinates between the two is given by\footnote{This can best be seen in the embedding space, where we think of AdS$_{d+1}$ as the surface $-X_0^2 - X_{d+1}^2 + \vec{X}^2 = - L^2$ in $\mathbb{R}^{2,d}$. The global AdS coordinates set $X_0^2 + X_{d+1}^2 = L^2 (1 + \tilde{r}^2)$, $|\vec{X}|^2 = L^2 \tilde{r}^2$. and $\tilde{t}$ is the angle along the $X_{0,1}$ circle. The global dS slicing sets $X_0=L \coth(z/L)$, $\vec{X}^2-X_{d+1}^2 = L^2/\sinh^2(z/L)$. The second equation is now solved by taking $X_{d+1} = L\sinh(\tau/L)/\sinh(z/L)$ and $|\vec{X}| =L\cosh(\tau/L)/\sinh(z/L)$. From this split the change of coordinates can easily be read off. We can also solve the second equation by picking out a direction in the $1,..,d$ directions, say $X_1$, and then taking $X_{d+1} = \sqrt{L^2-r^2} \sinh(t/L)/\sinh(z/L)$, $X_1 = \sqrt{L^2-r^2} \cosh(t/L)/\sinh(z/L)$, $X_{2,\ldots} = r/\sinh(z/L)$. This gives the part of global AdS$_{d+1}$ foliated by slices of the static patch of dS$_d$.}
\be
\tilde{r}= L \frac{\cosh(\tau/L)}{\sinh(z/L)}\,, \quad \tan\left(\frac{\tilde{t}}{L}\right) = \frac{\cosh(z/L)}{\sinh(\tau/L)}\,.
\ee

\begin{figure}[t]
\includegraphics[width=7cm]{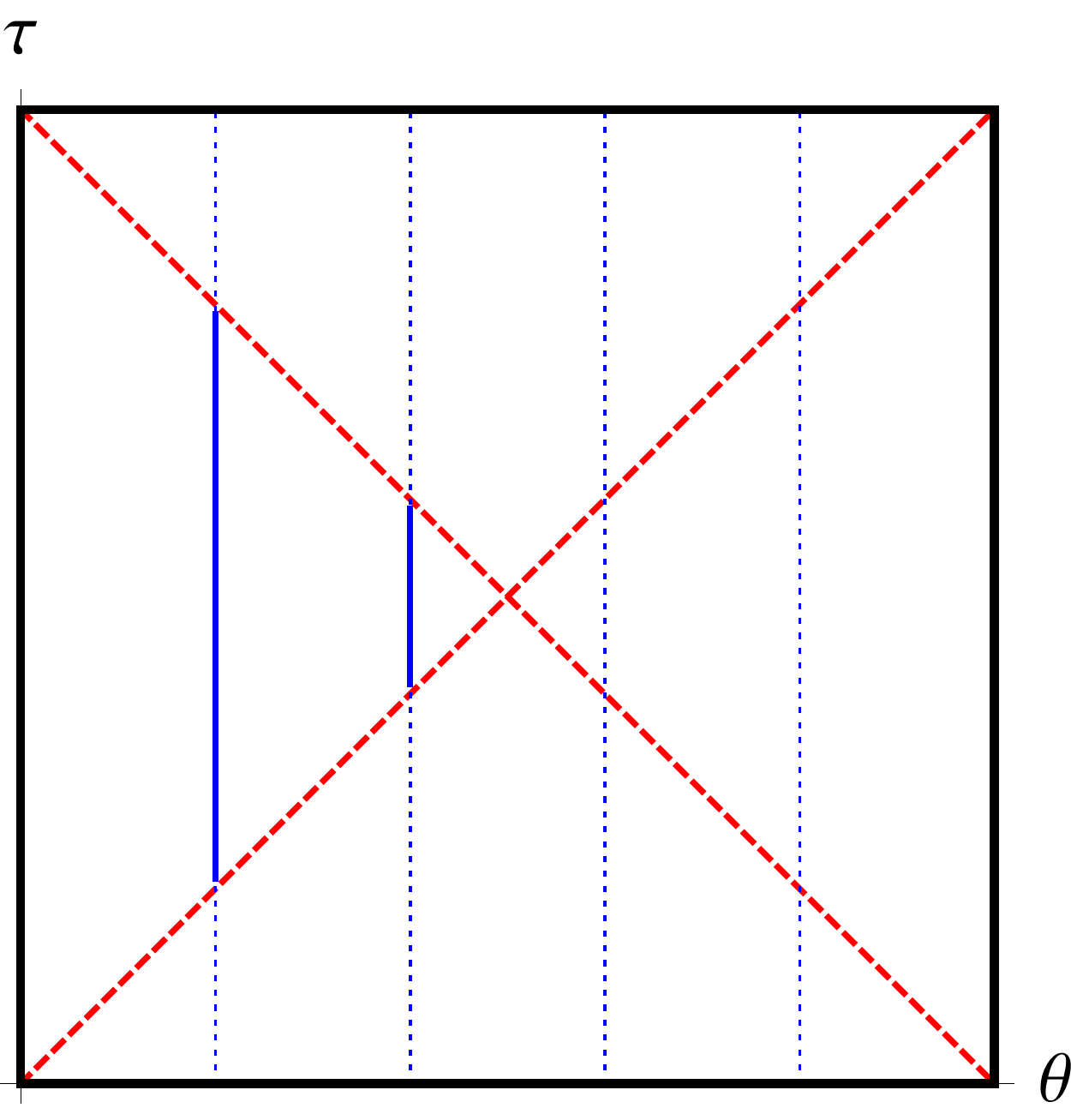}
\caption{Penrose diagram of de Sitter space. The global coordinates \eqref{globalds2} cover the whole square, with $\theta$ being the horizontal axis and time the vertical. The $d-2$ sphere is suppressed. The (red) dashed diagonal lines are the horizons of the static patch, (blue) dotted vertical lines are worldlines of particles sitting at different constant values of $\theta$. The solid blue lines are the part of these worldlines inside the static patch. From the point of view of the observer on the south pole these particles come out of the horizon, reach a minimum value of $r$, and then fall back into the horizon. The two rightmost worldlines are entirely outside the static patch of the south pole observer. A string connecting, say, the first and last worldline encodes an entangled ``Hawking" pair with particle and antiparticle being on opposite sides of the cosmological horizon.}
\label{fig:dspenrose}
\end{figure}

To understand how this same pair looks to the observer at the center of the static patch, we need to change coordinates on the slice from global coordinates \eqref{globalds2} to the static patch \eqref{staticds},
\be
	\frac{r}{L} = \cosh \left ( \frac{\tau}{L} \right ) \sin(\theta)\,,
	\quad
	\tanh \left ( \frac{t}{L} \right ) =
		\frac{\tanh \left ( \tau/L \right )}{ \cos(\theta)}\,.
\ee
Since $0 \leq r \leq L$, this change of coordinates only gives $r$ inside the static patch associated with the south pole of global dS for $ 0 \leq \theta \leq \pi/2$. The region $\pi/2 \leq \theta \leq \pi$ ends up behind the horizon in the static patch associated with an observer at the north pole. As we constructed string solutions with the endpoints at arbitrary $\theta_q$ and $\theta_{\bar{q}}$, we can put either both, neither, or one of the particles inside the static patch. For a given $0\leq \theta_q \leq \pi/2$ the quark follows a trajectory in the static patch given by
\be
\cot(\theta_q) =   \cosh(t/L) \frac{\sqrt{L^2-r^2}}{r}\,.
\ee
The quark comes out of the horizon at $t \rightarrow -\infty$, reaches a minimal value of $r$ set by $\theta_q$ at $t=0$, and then falls back into the horizon as $t\rightarrow \infty$. As before for the case of a $q$-$\bar{q}$ Hawking pair on opposite sides of an AdS black brane, if one quark is in the static patch, the string connecting the quark and anti-quark in the bulk will have to cross the horizon of the patch as well. In this way, they inherit a non-traversable wormhole on the worldsheet. In Figure \ref{fig:dspenrose}, we show the Penrose diagram for the global dS on each slice in order to illustrate the change of coordinates from global to static coordinates.

One quantitative measure of the entanglement of the $q$-$\bar{q}$ pair is the position-space entanglement entropy, wherein we trace over degrees of freedom in a region containing the $\bar{q}$. For the string,~\eqref{solution} with $\alpha=0$, hanging from the north pole of the boundary sphere down to the south pole, the method of~\cite{Casini:2011kv}, which relates the EE of a spherical region centered on the north pole to a thermal entropy on hyperbolic space, can be suitably generalized~\cite{Jensen:2013lxa} in the presence of the $q$-$\bar{q}$ pair. This in turn can be understood as a spherical EE, {\it{i.e}} for a sphere centered on the $q$, for mSYM on dS$_4$. The end result is that the EE for mSYM with the $q$-$\bar{q}$ pair is~\cite{Jensen:2013ora}
\be
\label{E:SEE}
S_{EE} = S_{\mathcal{N}=4} + \frac{\sqrt{\lambda}}{3} + \hdots\,,
\ee
where $S_{\mathcal{N}=4}$ is the EE in the absence of the $q$-$\bar{q}$ pair and the dots indicate subleading corrections at large $N$ and $\lambda$. The exact result for $S_{EE}$ has also been obtained~\cite{Lewkowycz:2013laa} by properly translating the exact result for a circular Wilson loop, which was previously calculated via localization, and of course agrees with~\eqref{E:SEE} in the usual $N\gg \lambda\gg 1 $ limits (with $\ln N \ll \sqrt{\lambda}$).

However, the methods described above~\cite{Casini:2011kv,Jensen:2013lxa} fail for any string~\eqref{solution} with $\alpha \neq 0$. The problem is that the quark is no longer static in the conformally related hyperbolic space, and so there is not a sensible definition of a thermal entropy. Although, the adaptation of generalized gravitational entropy~\cite{Lewkowycz:2013nqa} to probe objects~\cite{Karch:2014ufa}, in this case a probe fundamental string, should work in this instance.

\section{Conclusions}
\label{S:conclude}

We have demonstrated that the holographic dual of entangled pairs in mSYM on curved backgrounds and/or coupled to dynamical gravity is given by a classical string worldsheet whose worldvolume exhibits an ER bridge. So at least in these examples the recently proposed ER=EPR relation of \cite{Maldacena:2013xja} should be read as an equivalence. The entanglement of the quark-antiquark pair and the worldsheet with a wormhole are two different mathematical descriptions for one and the same mathematical reality. This is not quite the way it was originally phrased by the authors, but it seems to be where our holographic studies point us to even after including the effects of dynamical gravity.

In this regard, let us reiterate a point made in the Introduction on the example of entangled black holes that was used in \cite{Maldacena:2013xja} when they introduced the ER=EPR conjecture. Our standard understanding of a single black hole is that one can talk about this quantum state using two different languages. It can be viewed as a collection of microscopic degrees of freedom with an $e^S$ fold degeneracy, or it can be viewed as a classical object described by the black hole geometry with an entropy given by its horizon area. This philosophy underlies for example the famous Strominger-Vafa entropy count for supersymmetric black holes \cite{Strominger:1996sh}. In that case, the microscopic degrees of freedom are a collection of D-branes and strings wrapping an internal torus so they are effectively pointlike in the non-compact dimensions. For large brane and string charges, there is an exponentially large degeneracy of states of the wrapped branes/strings. The resulting entropy agrees with the Bekenstein-Hawking entropy of the black hole with the same charges. The two descriptions are dual to each other, in the sense that they calculate the same observable, a thermal entropy, of the same system in very different limits. The microscopic picture is valid in the weak coupling limit of the field theory on the brane/string intersection, while the macroscopic description in terms of a black hole holds at strong coupling.

Now we can easily imagine taking a doubled version of this theory, which has two sets of the wrapped branes/strings, and studying an entangled state between the two copies. In the microscopic picture, this is all there is to say. We study an entangled configuration in a theory which has two sets of wrapped D-branes. In the spacetime description that emerges at strong coupling, the dual macroscopic geometry is that of a wormhole. This is an example of ER=EPR, as the two descriptions reflect the same physical reality. In this work we have seen that the same is true for EPR pairs made of single $q$-$\bar{q}$ pairs, where the dual wormhole lives on the worldsheet of the ``fluxtube.''

\section*{Acknowledgments}
\noindent We would like to thank M.~Rangamani for useful correspondence. We would also like to thank Stefan Janiszewski for useful conversations. The work of KJ was supported in part by National Science Foundation under grant PHY-0969739. The work of AK and BR was supported, in part, by the US Department of Energy under grant number DE-FG02-96ER40956.

\bibliography{hawkingstring}

\end{document}